\documentclass[fleqn,usenatbib]{mnras}

\usepackage{newtxtext,newtxmath}

\usepackage[T1]{fontenc}

\DeclareRobustCommand{\VAN}[3]{#2}
\let\VANthebibliography\thebibliography
\def\thebibliography{\DeclareRobustCommand{\VAN}[3]{##3}\VANthebibliography}

\usepackage{graphicx}	
\usepackage{amsmath}	

\title[Energisation in collapsing magnetic traps]{A detailed investigation of particle energisation mechanisms in models
of collapsing magnetic traps}

\author[K. Mowbray, T. Neukirch and J. Threlfall]{
Kate Mowbray,$^{1}$\thanks{E-mail: jm380@st-andrews.ac.uk}
Thomas Neukirch,$^{1}$
James Threlfall$^{2}$
\\
$^{1}$School of Mathematics and Statistics,
University of St Andrews, St Andrews KY16 9SS, UK\\
$^{2}$School of Design and Informatics, Abertay University, Bell Street, Dundee DD1 1HG, UK\\
}

\date{Accepted XXX. Received YYY; in original form ZZZ}

\pubyear{2024}

\begin{document}
\label{firstpage}
\pagerange{\pageref{firstpage}--\pageref{lastpage}}
\maketitle

\begin{abstract}
In this paper we provide a detailed investigation of the energisation processes in two-dimensional, two and a half-dimensional and three-dimensional collapsing magnetic 
trap models. 
Using kinematic magnetohydrodynamic models of collapsing magnetic traps, we examine the importance of Fermi acceleration in comparison with betatron acceleration in these models. 
We extend previous work by investigating particle orbits in two-dimensional models without and with a guide field component and from full three-dimensional models. We compare the outcomes
for the different models and how they depend on the chosen initial conditions.
%
%
While in
the literature betatron acceleration has been emphasised as the major mechanism for particle energisation in collapsing magnetic traps, we find that Fermi acceleration can play a significant role as well for particle orbits with suitable initial conditions. 
\end{abstract}




\section{Introduction}
\label{sec:intro}


Magnetic activity processes in astrophysical plasmas
often involve the generation and transport of non-thermal particle populations \citep[e.g][]{benz02,
zharkova:etal11,birn:etal12,klein:dalla19,hoshino22,oka:etal23,guo:etal24}.
Well-known examples from within
our solar system are solar flares \citep[e.g.][]{krucker:etal08,zharkova:etal11,cargill:etal12,benz17} and
magnetospheric substorms 
\citep[e.g.][]{birn:etal12,oka:etal23}.
In these examples, magnetic reconnection allows the stored magnetic energy to be converted into thermal energy, kinetic energy associated with bulk flows and the generation of non-thermal particle populations. In solar flares, we observe the consequences of this process across the electromagnetic spectrum, for example in the form of radio emission \citep[e.g.][]{pick:vilmer08}, or as hard X-ray (and sometimes gamma ray) emission. X-ray sources are typically located at the footpoints and at the tops of magnetic field loops \citep[e.g.][]{krucker:etal08}. 
Although the motivation for the investigation presented in this paper is particle energisation and transport in solar flares, 
the models and the results we present could be applicable, with appropriate adjustments, 
to other space and astrophysical systems \citep[e.g.][]{birn:etal12,hoshino22,oka:etal23}.  

%
Various mechanisms have been proposed for particle energisation in solar flares. This includes acceleration directly associated with the magnetic reconnection process, either via the parallel electric field inside the diffusion region \citep[e.g.][]{zharkova-gordovskyy2004,zharkova-gordovskyy2005,wood:neukirch05,dalla-browning2005, dalla-browning2006,
dalla-browning2008,stanier-et-al2012,
threlfall:etal16a,threlfall:etal17,borissov:etal17,
borissov:etal20,gordovskyy:etal14,gordovskyy:etal20,
gordovskyy:etal23,pallister:etal19,pallister:etal21},
or with additional processes operating in the reconnection outflow region \citep[e.g][]{dahlin:etal14,dahlin:etal15,dahlin:etal17,drake:etal19}. The effect of multiple reconnection sites in a turbulent plasma state on particle acceleration has also been investigated \citep[e.g.][]{isliker:etal19, lazarian:etal12,turkmani:etal05,vlahos:etal04, vlahos:isliker19}. Other mechanisms that have been suggested are acceleration at a reconnection outflow induced termination shock \citep[e.g][]{cargill91,tsuneta:naito98,selkowitz:blackman04,miteva:mann07,mann:etal09,chen:etal15} and stochastic acceleration by the interaction of the particles with waves \citep[e.g][]{miller:etal97} or plasma turbulence \citep[e.g][]{liu:etal08}.
We remark that a combination of different mechanisms could contribute to particle acceleration for the same eruptive event. 

%
In the current paper we shall focus on another mechanism that has been suggested to contribute 
to particle energisation in flares, namely collapsing magnetic traps 
\citep[][]{somov:kosugi97}, which we will henceforth refer to as CMTs. The idea behind the CMT acceleration mechanism is that particles 
are trapped within a magnetic field configuration that is initially stretched. As the field rapidly relaxes over time, particles can gain
energy through both betatron acceleration, resulting from the increase in magnetic field strength,
and from Fermi acceleration, which in simple models of CMTs is a consequence of the shortening of the magnetic field lines \citep[the actual acceleration mechanism is more complicated in more sophisticated CMT models, e.g.][]{giuliani:etal05,eradat_oskoui:etal14}.
It is worthwhile to point out that similar energisation mechanisms have been associated with the magnetic field dipolarisation process in magnetospheric substorms 
\citep[e.g][]{birn:etal97,birn:etal98,birn:etal04,fu-et-al2013,khotyaintsev-et-al2011,artemyev14}.


The features that make CMTs encouraging as a contributor to particle energisation are that (i) a significant number of energised particles can remain trapped around the loop top region, hence providing a possible explanation for hard X-ray loop top sources as suggested by \citet{somov:kosugi97}, and (ii) a CMT can also enclose a relatively large volume of space and as a result, can potentially energise a large number of particles.

Previous investigations of the acceleration processes in CMTs range from those based on relatively simple magnetic field models
\citep[e.g.][]{kovalev:somov02,somov:bogachev03,kovalev:somov03a,aschwanden04,bogachev:somov07,bogachev:somov09,shabalin:etal22ApJ}
over somewhat more sophisticated analytical kinematic magnetohydrodynamic (MHD) models \citep[e.g][]
{giuliani:etal05,grady:neukirch09,minoshima:etal10,minoshima:etal11,grady:etal12,
eradat_oskoui:etal14,eradat_oskoui:neukirch14,borissov:etal16} to investigations based on numerical MHD simulations \citep[e.g.][]{karlicky:barta06}.
%



In this paper we shall use the kinematic MHD modelling framework 
first described for 2D CMT models in \citet{giuliani:etal05}, 
and
extended to 2.5D and 3D by \citet{grady:neukirch09}, to shed light on two specific aspects of CMT energisation processes, namely a) the relative importance of betatron acceleration and Fermi acceleration in CMTs, and b) whether and how the
energisation processes change from 2D to 2.5D and 3D models.


The reasons for this are as follows: 
firstly, when one considers 2D magnetic field models with curved magnetic field lines, as for example in \citet{giuliani:etal05}, it becomes impossible to completely 
disentangle betatron acceleration and Fermi acceleration, 
as is possible in quasi-one-dimensional models \citep[e.g][]{somov04,bogachev:somov05}. Furthermore,
as has been pointed out first by \citet{giuliani:etal05} 
and investigated in more detail by \citet{eradat_oskoui:etal14}, the particle energisation in the direction
parallel to the magnetic field is actually linked to the curvature of the magnetic field lines and hence takes place
where the field line curvature is large, i.e. usually at the loop tops of the models. Hence, linking Fermi acceleration
to the shortening of magnetic fields lines, i.e. the distance along a magnetic field line between
magnetic mirror points decreasing,
is misleading. We point out that
the 
acceleration
mechanism linked to field line curvature has also been proposed to operate at a microscopic level  for particle acceleration 
associated with
magnetic reconnection
in collisionless plasmas \citep[see e.g][]{drake:etal19}.
Secondly, previous studies based on 2D CMT models  \citep[e.g.][]{giuliani:etal05,grady:etal12} have identified betatron acceleration as the main contributor to energy gains of up to about 40 times the 
initial energy. Such gains are directly correlated with the increase of magnetic field 
strength experienced by the particles along their orbit. However, 
these energy gains are only found for particle orbits with specific initial conditions, 
with these being orbits that
are initially located in the weakest regions of the field close to the centre of the CMT and thus experience
the largest increase in magnetic field strength 
over time.
It is therefore important to investigate the contribution of Fermi acceleration in CMTs and how it depends on the initial conditions of particle orbits. 

A closely related point is 
the finding by \citet{birn:etal17} that when studying particle energisation in a 3D MHD simulation of an erupting sheared magnetic arcade, the strong guide field of the configuration limited the magnetic field compression factor and hence the effect of betatron acceleration to much lower values than found in the 2D kinematic models. 
In order to investigate these findings in more detail and to assess the relative importance of betatron and Fermi acceleration in similar configurations, one needs CMT models that are 2.5D or 3D, i.e. have the ability to have a magnetic field with three non-vanishing components. 
\citet{grady:neukirch09} have extended the theoretical framework for kinematic MHD CMT models to 2.5D and 3D,
but a systematic investigation of particle orbits in
such models has yet to be carried out and it is one
of the aims of this paper to provide an initial
survey of the energisation processes in higher dimensional
CMT models.
The paper is organised as follows. In
section 2 we provide a summary of the theoretical framework underpinning the CMT models we use, 
and an overview of the numerical method used to calculate particle trajectories and energies. In section 3 we will present our results. The first set of results that we present will assess the impact of both betatron and Fermi acceleration in the 2D CMT model first presented in \citet{giuliani:etal05}. Following this, we will discuss the importance of these energisation terms in both the 2.5D and an adjusted 3D model from \citet{grady:neukirch09}, in particular focussing on how the relative importance of these processes vary in more complicated field configurations like a twisted or sheared magnetic field. In section 4 we summarise our findings and assess their 
implications,
as well as
discussing
how the CMT model could be improved to make it more realistic.

\section{Basic Theory and Numerical Method}
\label{sec:basic}

\subsection{CMT Theory Overview}
\label{ssec:CMTtheory}

 To generate the CMT fields, we
 start from
 the 2D model detailed in \citet{giuliani:etal05}
 and the 2.5D and 3D models detailed in 
 \citet{grady:neukirch09}.
 To find the relevant electric and magnetic fields for our CMT models, we use a prescribed plasma velocity 
 %
%
 field 
 to solve the ideal kinematic MHD equations given by:

\begin{eqnarray}
    \label{eqn:ohmslaw}
    \textbf{E} + \textbf{V} \times \textbf{B} &=& \textbf{0}, \\
    \label{eqn:faradayslaw}
    \frac{\partial \textbf{B}}{\partial t} &=& -\nabla \times \textbf{E}, \\
    \label{eqn:solenoidalcon}
    \nabla \cdot \textbf{B} &=& 0,
\end{eqnarray}


\begin{table}
%
%
  \caption{Typical normalising values used for the CMT models in this paper \citep[following][]{giuliani:etal05}}
    \centering
    \begin{tabular}{l|c|l}
     Length scale     &   L            & 10 Mm \\
     Time scale       & $t_{scl}$      & 100 s \\
     Speed            & $V_{scl}$       & $10^{5}$ m/s \\
     Magnetic field   &   $B_{scl}$    & 0.01 T \\
    Larmor frequency & $\Omega_{scl}$  &  $1.7\cdot 10^{9}$ s$^{-1}$ 
    \end{tabular}
    \label{tab:scales}
\end{table}

where $\textbf{B}$ is the magnetic field, $\textbf{E}$ is the electric field and $\textbf{V}$ is the prescibed velocity
%
%
field, 
defined 
implicitly
by a space- and time-dependent coordinate transformation. This coordinate transformation acts on the specified field $\textbf{B}_{\infty}$, the field to which the CMT relaxes to as $t$ tends to $\infty$. In 2D and 2.5D we specify this field using a flux function and in 3D we specify the field itself. The only constraint we have on our field $\textbf{B}_{\infty}$, is that it must satisfy the solenoidal condition. The field can be specified at any finite time using the coordinate transformation alongside $\textbf{B}_{\infty}$. For both 2D and 3D versions of the model, this method will ensure that fields satisfy the ideal kinematic MHD equations at all times. Details of this method using the flux function for 2D and 2.5D are given in
\citet{giuliani:etal05}
and for 3D using Euler potentials in the appendix of 
\citet{grady:neukirch09}.

The model uses the 
normalisation values 
provided
in Table \ref{tab:scales}. To allow for consistent notation between 2D, 2.5D and 3D models, we have the $y$ coordinate point outwards from the solar surface for all models, with $y=0$ corresponding to the 
%
%
lower boundary of our calculational domain,
with heights above this boundary being associated with the solar 
corona.
The $x$ direction runs parallel to the solar surface and the $z$ direction is 
the invariant direction for the 2.5D model.  In the following we assume that all quantities are normalised, for example all coordinates are measured in units of $L$, and so on.

The 2D model 
we use was first detailed in
\citet{giuliani:etal05}
and uses the following coordinate transformation and flux function:

\begin{eqnarray}
\label{eqn:2dxinf}
x_{\infty} &=& x, \\
\label{eqn:2dyinf}
y_{\infty} &=& (at)^{b}\ln\left[1+\frac{y}{at^{b}}\right]\left\{\frac{1+\tanh [(y-L_{v})a_{1}]}{2}\right\} \notag \\ 
&& +\left\{\frac{1-\tanh [(y-L_{v})a_{1}]}{2}\right\}y, \\
\label{eqn:2dainf}
A_{\infty} &=& c_{1}\left[\arctan \left( \frac{y_{0}+d}{x_{0}+1/2} \right) - \arctan \left( \frac{y_{0}+d}{x_{0}-1/2}\right) \right].
\end{eqnarray}
The functions $x_{\infty}$ and $y_{\infty}$ are functions of space and/or time which describe the coordinate transformation. These equations make use of the constant values $a$, $b$, $L_v$, $a_1$, $c_1$ and $d$ which will be explained later in this section. This flux function and coordinate transformation can be used to obtain the magnetic field at finite times using:

\begin{eqnarray}
    \label{eqn:Bxin2D}
    B_x &=& \frac{\partial A_{\infty}}{\partial y} = 
    \frac{\partial A_{\infty}}{\partial y_{\infty}} \frac{\partial y_{\infty}}{\partial y} +
    \frac{\partial A_{\infty}}{\partial x_{\infty}} \frac{\partial x_{\infty}}{\partial y} \\
    \label{eqn:Byin2D}
    B_y &=& -\frac{\partial A_{\infty}}{\partial x} = 
    -\left(\frac{\partial A_{\infty}}{\partial y_{\infty}} \frac{\partial y_{\infty}}{\partial x} +
    \frac{\partial A_{\infty}}{\partial x_{\infty}} \frac{\partial x_{\infty}}{\partial x}\right).
\end{eqnarray}

This choice of $A_{\infty}$ will see the CMT collapse towards a potential field generated by two magnetic sources of opposite charge placed at $(1/2,-d)$ and 
$(-1/2,-d)$ in normalised coordinates. For our model we set $d=1$. The parameter $c_{1}$ specifies the strength of the charges and for our simulations takes the value $-1.5 \times 10^{5}$ Tm. This yields a final field with symmetrical collapsed loops pinching in towards narrow footpoints at the solar surface.

For our solar application of the model, $y=0$ is the 
%
%
lower boundary of our calculational domain,
with heights above this point being associated with the solar 
corona.
The increased density of the 
solar atmosphere below the corona
means that we consider particles to be `lost' from the system if they pass below $y=0$.
%
%

The 
transformation specified in Eqn.\ (\ref{eqn:2dyinf})
leads to the magnetic field
configuration 
becoming
stretched in the $y$-direction for earlier times 
in the model,
with this stretching taking place in the region above $y = L_v$. We take $L_v = 1$. The steepness of the transition between regions where the transformation acts to stretch field lines and where it does not is controlled by $a_1$ (= $0.9$). The time-dependent collapse of the CMT is governed by the $(at)^b$ term found in Eqn.\ (\ref{eqn:2dyinf}), where we set $a = 0.4$ and $b = 1.0$.
As 
$t$ tends to $\infty$, $y_{\infty}$ will approach $y$, so that the coordinate transformation $\textbf{X}_{\infty} = (x_{\infty},y_{\infty})$ reduces to the identity transformation, ensuring that the long term state of the field will indeed be the potential field generated by $A_{\infty}$.  

In 
\citet{grady:neukirch09}
this model is extended to 2.5D by specifying a constant $B_{z final}$ which describes the strength of the $z$ component of the field for large times. At any given time, the component $B_{z}$ will be given by $\frac{\partial y_{\infty}}{\partial y} B_{z final}$. The coordinate transformation eventually reduces to the identity, ensuring that the $z$ component of the field will eventually tend towards $B_{z final}$. When extending to 2.5D, the long term state of the field will now be the potential 
%
%
field
generated by lines of magnetic sources at $x=\pm 0.5$ and at $y = -d $, with both lines extending in the $z$ direction. The field will relax to the same final field as the 2D case when projected onto the $x$-$y$-plane. The constant $B_{z final}$ specifies the strength of the $B_{z}$ component, with this component tending towards $B_{z final}$ with increasing time. 

Using a 3D CMT model gives us the freedom to generate a more realistic field configuration. We use a variation of the 3D model described in 
\citet{grady:neukirch09},
which uses a coordinate transformation that 
%
%
is able to both stretch and twist field lines 
in the initial configuration, leading to a field that untwists 
as it collapses to the potential field. 
This model uses the same stretching term of $y_{\infty}$ as the 2D model written in Eqn. (\ref{eqn:2dyinf}), alongside the following coordinate transformations for $x$ and $z$:

\begin{eqnarray}
\label{eqn:3dxinf}
x_{\infty} &=& x - \delta\,[y_{\infty}(y,t)-y]\frac{z}{
D}, \\
\label{eqn:3dzinf}
z_{\infty} &=& z 
+ \delta\,[y_{\infty}(y,t)-y]\frac{x}{
D}, 
\end{eqnarray}
where $D=a_{3D}^{2}+x^{2}+[a_{y}(y-y_{h})]^{2}+z^{2}$,
coupled with the final field

\begin{eqnarray}
\label{eqn:3dbinf}
\textbf{B}_{\infty} &=& \bar{c}_{1}\frac{\left[ \left(x_{0}+\frac{L}{2}\right)\textbf{e}_{x}+(y_{0}+d)\textbf{e}_{y}+z_{0}\textbf{e}_{z}\right]}{\left[(x_{0}+L/2)^{2}+(y_{0}+d)^{2}+z_{0}^{2}\right]^{3/2}} \notag \\
&&-\bar{c}_{1}\frac{\left[ \left(x_{0}-\frac{L}{2}\right)\textbf{e}_{x}+(y_{0}+d)\textbf{e}_{y}+z_{0}\textbf{e}_{z}\right]}{\left[(x_{0}-L/2)^{2}+(y_{0}+d)^{2}+z_{0}^{2}\right]^{3/2}}, 
\end{eqnarray} 

where $\bar{c}_{1} = -1.5\times 10^{5}Tm^{2}$, $\delta$ is a parameter controlling the strength of the twist in the field and $a_{3D}$ (= 1.0) is a constant which prevents the 
%
%
denominator $D$
in Eq.~(\ref{eqn:3dxinf}) and Eq.~(\ref{eqn:3dzinf}) from becoming zero. The parameters $a_y$ and $y_h$, which present a departure from the model in \citet{grady:neukirch09}, are discussed below.\\

%
%
A closer inspection of
the 3D model detailed in 
\citet{grady:neukirch09}
%
%
identified 
a problem with this model. The most extended field lines
%
%
experience
a very strong twist, accompanied by a stronger field strength, some way up the loop legs. This gave rise to unrealistic particle orbits. Examining the variation of the field strength with $y$ showed that after the appearance of a local maximum, the field strength tended towards a positive constant value, implying that a magnetic field formed by two sources would induce a finite field strength
%
%
as $y \to \infty$.
%
%
%

%
The problem in the \citet{grady:neukirch09} model is
caused by the $y_{\infty}(y,t)-y$ dependence of their transformation. which is similar to
our Eqs.\ (\ref{eqn:3dxinf}) 
and (\ref{eqn:3dzinf}), but does not include any $y$-dependent terms in the denominator.
This
leads to the twist becoming stronger with 
increasing $y$ for finite times. Our modification of the transformation rectifies 
the problem by introducing the $a^2_{y}(y-y_{h})^{2}$ term in the denominator of 
both $x_{\infty}$ and $z_{\infty}$. 
This additional term ensures that the twist in the 
field tends to zero for large values of $y$. 
This modification leads to the introduction of two new parameters: $a_{y}$ which 
controls how quickly the twist is reduced for 
increasing $y$, and $y_{h}$ which 
specifies the height at which the reduction first begins to take effect.

\subsection{Particle Orbits}
\label{ssec:particleorbits}

We investigate the motion and energisation of particles in these fields using test particle 
%
%
calculations.
To 
%
%
determine
the particle orbits and their energies we make use of the relativistic guiding centre 
%
%
equations detailed in
\citet{northrop63}.
\begin{eqnarray}
\label{eqn:dupardt}
\mbox{\hspace{-0.6cm}} \frac{du_{\parallel}}{dt} = \frac{d}{dt}(\gamma v_{\parallel}) &=& \gamma \mathbf{u}_{E} \cdot \frac{d\mathbf{b}}{dt}+\Omega_{scl}t_{scl}E_{\parallel}-\frac{\mu_{r}}{\gamma}\frac{\partial B^{*}}{\partial s}, \\
\label{eqn:drperpdt}
\dot{\mathbf{R}}_{\perp} &=& \mathbf{u}_{E} + \frac{\mathbf{b}}{B^{**}} \times \\
&& \left\{ \frac{1}{\Omega_{scl}t_{scl}} \left[\frac{\mu_{r}}{\gamma} \left(\nabla B^{*} + \frac{V_{scl}^{2}}{c^{2}}\mathbf{u}_{E}\frac{\partial B^{*}}{\partial t}\right) \right. \right. \notag \\
&& \left. \left. + u_{\parallel}\frac{d\mathbf{b}}{dt} + \gamma \frac{d\mathbf{u}_{E}}{dt}\right] +  
 \frac{V_{scl}^{2}}{c^{2}}\frac{u_{\parallel}}{\gamma}E_{\parallel}\mathbf{u}_{E}\right\}, \notag \\
\label{eqn:gammadefn}
\gamma &=& \sqrt{1+\frac{u_{\parallel}^{2}+u_{E}^2}{c^{2}}+\frac{2\mu_{r} B}{mc^{2}}}, \\
\label{eqn:mudefn}
\mu_{r} &=& \left(\frac{\gamma^{2}mv_{\perp}^{2}}{2B}\right), 
\end{eqnarray} 

where $\mu_{r}$ is the relativistic 
magnetic moment, 
%
%
%
$\mathbf{u}_{E} = \gamma \mathbf{v}_E$, where
$\mathbf{v}_E =
\frac{\mathbf{E} \times \mathbf{B}}{B^{2}}$ is the 
$\mathbf{E} \times \mathbf{B}$ drift, $\mathbf{b} = \mathbf{B}/B$ and $\dot{\mathbf{R}}_{\perp}$ is the perpendicular motion of the guiding centre across field lines resulting from drift terms, whilst $v_{\parallel}$ and $v_{\perp}$ are the parallel and perpendicular components of the particle's velocity to the magnetic field. Additionally, we have that:

\begin{equation}
    B^{*} = B\left(1-\frac{E_{\perp}^{2}}{c^{2}B^{2}}\right)^{1/2}, B^{**} = B\left(1-\frac{E_{\perp}^{2}}{c^{2}B^{2}}\right). \notag
\end{equation}

With the length scales and time scales set as described in Table \ref{tab:scales}, 
%
%
%
using 
the guiding centre approximation
%
%
is clearly 
justified,
with typical length scales being much larger than the gyroradius and the typical time scale being far greater than the time period of the gyrational motion. Our magnetic field models only cover the regions outside the reconnection region and will 
%
%
therefore not include any domains with parallel electric fields, or which contain magnetic null points. 
This means that the gyroradius of particles (which scales as $\frac{1}{B}$) will remain well below our typical length scale of $10^{7}$m, ensuring the validity of the guiding centre approximation. By staying in the ideal region, we can see from Equation (\ref{eqn:ohmslaw}) that the electric field and magnetic field will always be perpendicular to each other. Consequently, particles in our model will not be accelerated by a parallel electric field.  


\subsection{Numerical Method}
\label{ssec:nummeth} 
To solve the relativistic guiding centre equations numerically, we use a numerical code that has been used before by
various authors \citep[e.g.][]{giuliani:etal05,grady:etal12,threlfall:etal15a,threlfall:etal16a,threlfall:etal16b,threlfall:etal17,borissov:etal16}.
The code uses a fourth-order Runge-Kutta method (RK4) with a variable time step that responds to the error, calculated using an RK5 method.  

We specify the initial conditions for an orbit and 
calculate
the orbit either for the $100$s duration of the simulation, or up until the time that an orbit escapes the CMT, whichever is smaller. When setting the initial conditions, we specify the initial energy and pitch angle, with the magnetic moment being calculated from this input. Particles with the same initial energies and pitch angles can have varying magnetic moments if their orbits start in positions that have different field strengths.  


In addition to equations (\ref{eqn:dupardt}) - (\ref{eqn:mudefn}), we make use of the expression for $\frac{d \gamma}{d t}$ 
%
%
for diagnostic purposes:

 \begin{equation}
\label{eqn:dgammadt}
\frac{d\gamma}{dt} = \frac{V_{scl}^{2}}{c^{2}}\left[\Omega_{scl}t_{scl}\left(\dot{\mathbf{R}}_{\perp}+\frac{u_{\parallel}}{\gamma}\mathbf{b}\right) \cdot \mathbf{E} + \frac{\mu_{r}}{\gamma}\frac{\partial B^{*}}{\partial t}\right]. \\
 \end{equation}

This equation is not used 
%
%
in the calculation of
particle orbits, but 
%
%
it is
a useful indicator of the particular energisation processes at play by breaking down the individual terms that are responsible for energy increases/decreases. The particle orbit code 
%
%
%
%
calculates each of the terms in Equation (\ref{eqn:dgammadt}) to assess their relative importance for 
%
%
changing the particle energy
during 
an
orbit.

\section{Results}

We vary initial positions in the magnetic field to assess the impact of different energisation processes at particular times and positions in the CMT. 
In each model 
%
%
orbits
are initialised to 
%
%
represent the
significant variation 
%
%
caused by different
%
%
field line shapes.
This will range from field lines that start in a mostly collapsed state to closed field lines that are stretched to the extent shown for the most stretched field line in the top panel of Figure \ref{fig:2dflines}.

\subsection{Separating Fermi and betatron acceleration}

As discussed in Section \ref{ssec:CMTtheory}, particles will not be energised by a parallel electric field in our CMT simulations. This leaves Fermi acceleration and betatron acceleration as the processes responsible for particle energisation. Of the two, betatron acceleration is easier to quantify. 
Substituting the drift terms in Equation (\ref{eqn:drperpdt}) into Equation (\ref{eqn:dgammadt}), the two terms describing betatron energisation are 
\begin{equation}
\frac{\mu_{r}}{\gamma}\frac{\partial B^{*}}{\partial t} + \mathbf{E} \cdot \left( \frac{\mathbf{b}}{B^{**}} \times  \left[ \frac{\mu_r}{\gamma}\nabla B^*\right] \right). \nonumber
\end{equation}
For $B^{**} \approx  B $ (i.e. when the plasma velocity is much less than the speed of light), this can be rearranged to approximately give:
\begin{equation}
\frac{\mu_{r}}{\gamma}\left(\frac{\partial B}{\partial t} + \mathbf{u}_E \cdot \nabla B\right), \nonumber
\end{equation}
so that the term in brackets represents the change in the magnetic field strength at a position on the field line due to the collapse of the field. 

A key observation at this point is that while the $\textbf{E} \times \textbf{B}$ drift, 
%
%
represented here by 
$\textbf{u}_{E}$, is the lowest order drift term, it will not contribute towards the energisation of the particle since it is always perpendicular to the electric field. The interaction between the electric field and the higher order drift terms, (except for the $\nabla B$ drift which describes betatron acceleration), will be responsible for any energisation due to Fermi acceleration as discussed in
\citet{giuliani:etal05}, \citet{grady:etal12}, and
\citet{eradat_oskoui:etal14}.

In order to separate the 
contributions
of Fermi and betatron acceleration, we will need to keep track of the terms relating to the four-velocity of the particle appearing in Equation (\ref{eqn:gammadefn}), which we will call:
\begin{equation}
    u_{tot}^{2} = u_{\parallel}^{2} + u_{E}^{2} + \frac{2 \mu_r B}{m}.
    \label{eqn:fullvelocity}
\end{equation}
Note that for the total particle velocity ($v_{tot}$) :
%
%
\begin{equation}
    u_{tot}^2 = \gamma^2 v_{tot}^2 \approx 
                \gamma^2 (v_\parallel^2 + v_{\perp}^2)
\end{equation}
as long as $v_E^2 \ll v_{\parallel}^2 + v_{\perp}^2$. 

Two difficulties complicate the separation of parallel and perpendicular energy components associated with a particle orbit. The first is the continual conversion of parallel to perpendicular velocity (and vice-versa) due to particle mirroring. 
This conversion obfuscates the impact of either of the energisation processes. We identify the individual impacts of Fermi and betatron acceleration by tracking energy gains due to the simpler process, betatron acceleration. With this contribution accounted for, we will be able to conclude that any energy gains or losses not explained by the betatron effect are the result of Fermi acceleration or deceleration. Looking to Equation (\ref{eqn:fullvelocity}), we can see that $\mu_r B$ is an indicator of the particle energy relating to the magnetic field strength. 
This term will increase as the magnetic field evolves and the magnetic field strengthens as it relaxes to a compressed state. This is precisely the effect of betatron acceleration. It is distinct from particle mirroring upon approaching regions of stronger field on a field line, which does not affect particle energies. 

Our second difficulty is that we can 
%
%
%
%
use
$\mu_r B$ only as an approximate indicator of the perpendicular energy of the particle because, in relativistic 
%
%
dynamics,
a straightforward division of the total energy between parallel and perpendicular components is not possible. In relativistic 
theory,
the particle kinetic energy is:
\begin{equation}
    \label{eqn:relenergy}
    E_{k} = mc^{2}(\gamma - 1),
\end{equation}
%
%
and thus, unlike the non-relativistic kinetic energy, the total kinetic energy is not the
sum of the parallel and perpendicular kinetic energies due to the square root in the definition of $\gamma$. 
%
%
This 
prevents a striaghtfoward decomposition of the kinetic energy into components
parallel or perpendicular to magnetic field lines. However, for non-relativistic 
particles (with $u_{tot}^{2} \ll c^{2}$), 
%
%
taking the non-relatistic limit
by expanding 
the square root in Equation (\ref{eqn:gammadefn}) yields
\begin{equation}
    \label{eqn:relenergyapprox}
    E_{k} = mc^{2}\frac{u_{tot}^{2}}{2c^{2}} + 
%
%
       \mathcal{O}\left(\left[\frac{u_{tot}^{2}}{c^{2}}\right]^{2}\right) 
    \approx \frac{1}{2} m (v_\parallel^2 + v_\perp^2),
\end{equation}
where we have neglected the contribution of the 
$\mathbf{E}\times\mathbf{B}$-drift as discussed above.

Thus $\mu_r B$ is a first order approximation of the perpendicular energy of non-relativistic particles. This should remain a valid approximation for the particle energies measured here but may break down if particles gain enough energy to become relativistic. 
%
%

Equation (\ref{eqn:mudefn}) provides an expression for $\mu_r$ which, when substituted into $\mu_r B$ yields:


\begin{equation}
    \mu_r B = \frac{\gamma_{t_{0}}^{2}mv_{\perp t_{0}}^{2}}{2}\left(\frac{B}{B_{t_{0}}}\right),
    \label{eqn:muB}
\end{equation}

where terms subscripted with $t_{0}$ are values taken at the initial position and time. Since $\mu$ is an adiabatic invariant, the value that Equation (\ref{eqn:mudefn}) takes at $t=0$ will hold for all times.

The term outside of the brackets in Equation (\ref{eqn:muB}) remains constant for each particle during a simulation. Further it remains constant between a set of particles starting with the same initial energies and pitch angles. By the end of a simulation, the energy gains due to betatron acceleration will be approximately proportional to the ratio of final to initial field strengths. A scatter plot of the ratio of final to initial energies against the ratio of final to initial magnetic field strengths for a number of particles can show how well particle energisation will be explained by betatron acceleration alone in the CMT. For a system in which betatron acceleration is dominant, we anticipate a straight line with a positive slope, where an increase in the magnetic field strength for a particle orbit directly translates to an increase in energy for the particle. Once plotted, any significant deviation from the straight line will indicate that the Fermi effect plays a key role in energising some particles. Points lying above the line would indicate Fermi acceleration; point lying below the line would indicate Fermi deceleration.

Care is needed in measuring the magnetic field strength at the end of an orbit 
%
%
calculation. 
If we measure the field strength at the point in the orbit where the simulation completes, particle trajectories may finish at different points on a field line. Orbits finishing closer to their mirror points will be positioned within regions of much stronger magnetic field than those finishing close to the loop top, introducing a 
%
%
systematic
error. Instead we will 
measure the final field strengths and energies at the final time that the orbit passes 
the loop top in order to consistently compare the final field strengths for a group of particles. This is the only point on a field line that all orbits can be guaranteed to pass through, since particle trajectories will be reversed at various distances into the loop legs. Only by measuring on the final loop top pass can we ensure that we measure a comparable contribution of $\mu B$ to the particle energy for all of our orbits. For the field lines in the model, the collapse of the field will have slowed significantly by the end of the simulation and particles will be making loop top passes regularly. As a result, the particle energies should change slowly at this time; while we may be comparing energies at slightly different times the error in the final energies will be minimal.

\subsection{2D Results}
\label{ssec:2Dresults}

\begin{figure}
    \includegraphics[width=\columnwidth,trim={0.8cm 0.8cm 1.2cm 1.0cm},clip]{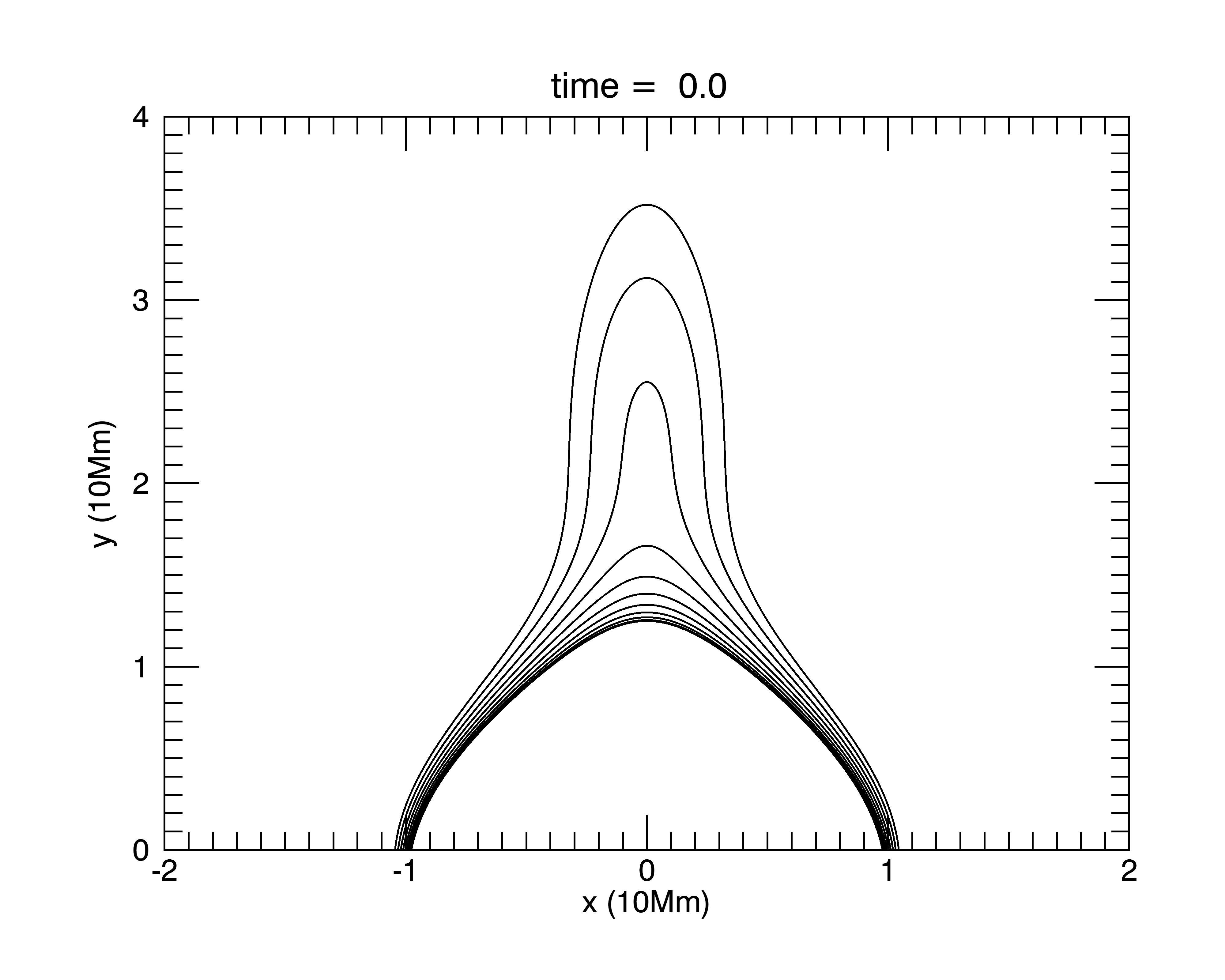}
    \includegraphics[width=\columnwidth,trim={0.8cm 0.8cm 1.2cm 0.0cm},clip]{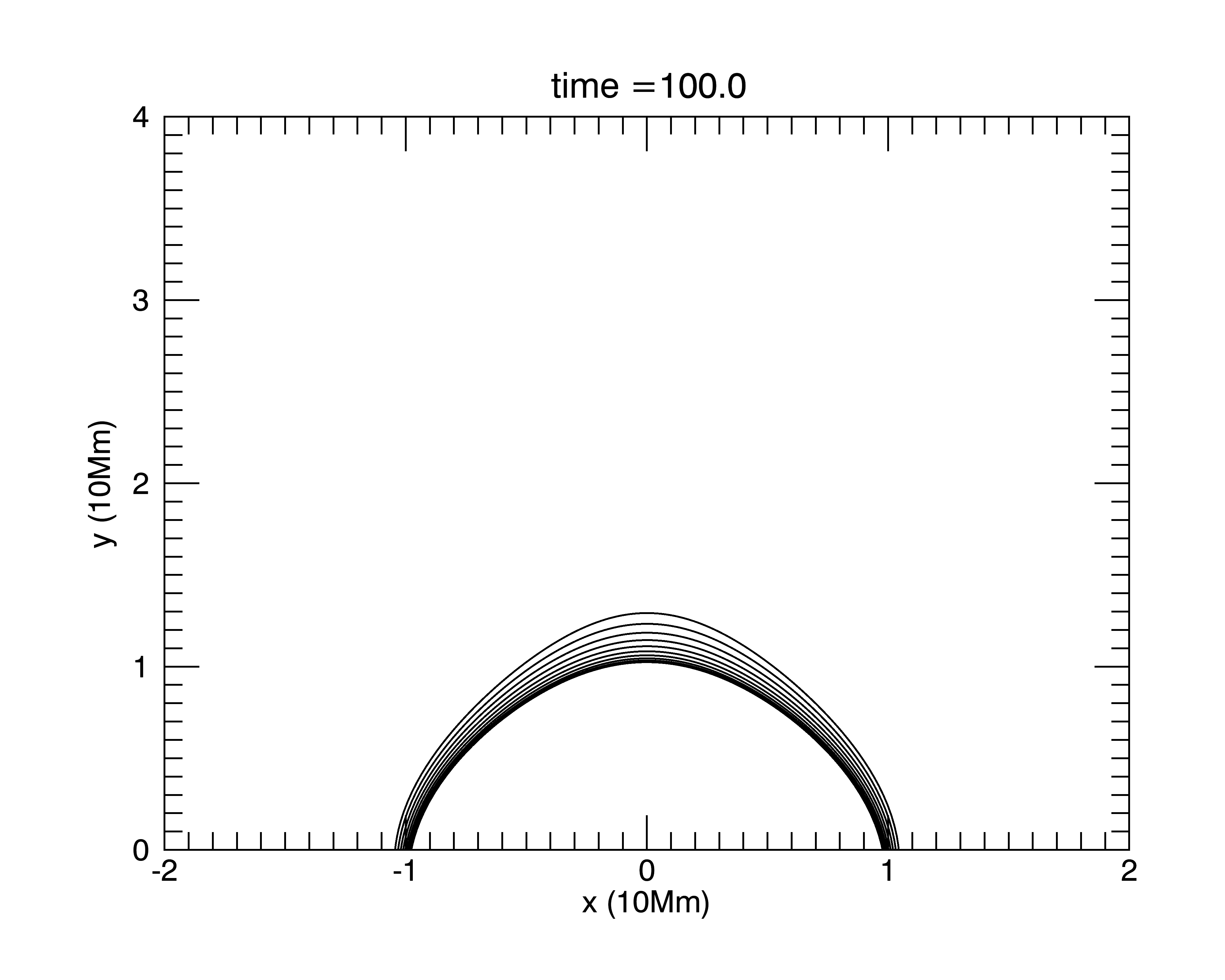}
    \caption{Illustration of the time evolution of selected field lines 
    in the 2D model at $t=0$s (top) and $t=100$s (bottom), with field lines traced from the initial positions used for particle orbits in Section \ref{ssec:2Dresults}.}
    \label{fig:2dflines}
\end{figure}

In our 2D model, to start orbits on a set of initial field lines, all stretched to varying degrees, we 
%
%
choose initial
positions lying on a straight line in $x$. 
%
%
Hence, all particle orbits start at a height of $y=1.25$, and we place the $121$ 
initial positions evenly between $x=-0.5$ and $x=0$. 
The top panel in Figure \ref{fig:2dflines} shows the initial states of field lines that these particle orbits start on. We use 
%
%
an initial particle 
energy 
of $5.5$keV, for better comparison with the results presented in 
\citet{grady:neukirch09} and \citet{grady:etal12},
%
%
who
also use this value (though the 
%
%
calculations
in these papers use a non-relativistic CMT model). This represents a relatively high initial energy when compared to the coronal thermal energy, indicating some amount of pre-acceleration of particles before they interact with the CMT. 
%
%
The initial pitch angle is set to $60^{\circ}$, so that the terms outside of the brackets in Equation (\ref{eqn:muB}) are identical for each particle orbit. This pitch angle is chosen to be high enough to keep particle trajectories trapped in the CMT for the full time that the simulation runs whilst also being low enough to avoid particles being constrained to only a narrow region in the field. This ensures that we compare particles which spend the same amount of time in the CMT. We will consider this the standard case for our initial conditions when adjusting parameters later.  

\begin{figure}
    \centering
    \includegraphics[width=\columnwidth, trim={0.8cm 0.8cm 1.32cm 1.55cm},clip]{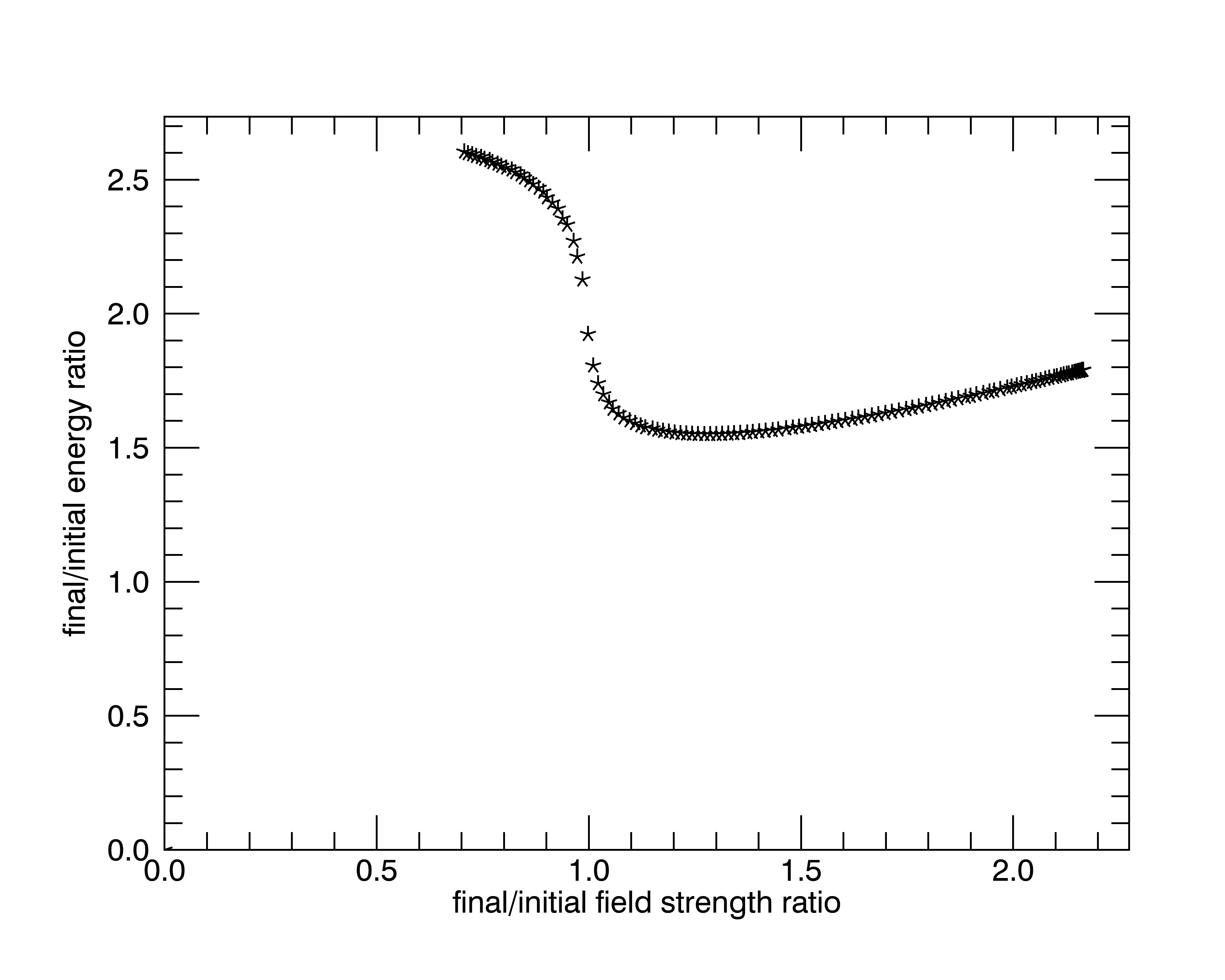}
    \caption{Ratio of final to initial energy against final to initial field strength for orbits in the 2D model. Initial energies are $5.5$keV and initial pitch angles are $60^{\circ}$.}
    \label{fig:2Dinx}
\end{figure}

The results for 
%
these 
initial conditions are presented in Figure \ref{fig:2Dinx}. It is important to note that even though some data points show a field strength ratio of less than one, none of the orbits experience betatron deceleration over the complete course of the simulation.
This is a consequence of our decision to measure the final field strength at the final loop top pass, meaning that field strength is always measured at the point on the field line where the field is weakest. The initial field strength is measured at the starting position of the orbit, which will sometimes be quite far into the loop legs, where the field is stronger. The horizontal axis in 
%
%
Figure \ref{fig:2Dinx} is an indicator of the expected energy gains resulting from betatron acceleration. Stronger initial fields lead to lower 
%
%
values of the magnetic moment $\mu$, resulting in more moderate energy gains from betatron acceleration across a particle orbit.  

Figure \ref{fig:2Dinx} displays three key features. The first of these is 
towards the right of the diagram, in the region where the ratio of the final to the initial field strength is about one or larger. 
This region follows a straight line quite closely, indicating that betatron acceleration is the dominant energisation process for particles on these orbits. Next is the region towards the top left, where the field strength ratio is less than approximately 0.9 and the energy ratio is between 2.3 and 2.7. Here the energy gains far outstrip those expected due to betatron acceleration alone. Particles on these orbits are experiencing significant energy gains as a result of Fermi acceleration. The third area of interest is found where the magnetic field ratio is between about 0.9 and 1.0, where the energy ratio reduces from around 2.3 to 1.5. This region, which shows a steep drop off in energies for increasing field strength ratio, connects the other two regions and will hereafter be referred to as the `transitional region'.

Figure \ref{fig:2Dinx} shows that the `transitional region' is key to understanding how the initial position of a particle orbit relates to the energy gain along that orbit. Orbits corresponding to points found at the top of this steep drop off start on field lines that initially lie just above the point in the field where the field strength is at its minimum. Orbits corresponding to points at the bottom of the `transitional region' start on field lines that initially lie just below the minimum. By inspecting orbits in the other two regions, those displayed on the right of the figure, which almost exclusively gain energy due to betatron acceleration initially start on the most collapsed field lines, whereas those displayed in the top left, which experience significant energy gains due to Fermi acceleration start on the most stretched field lines. This indicates the importance of collapsing field lines in energising particles due to Fermi acceleration and shows the particular importance of the minimum of the field strength in this process. Field lines collapse fastest at this minimum because it corresponds to the point at which the coordinate transformation $y_{\infty}$ stretches field lines the most. The field is weakest in this region because of this stretching. 
Additionally, because the plasma velocity is entirely perpendicular to the field at loop tops they are the locations where field lines move fastest during the collapse.

\begin{figure}
    \centering
    \includegraphics[width=0.99\columnwidth, trim={0.7cm 0.5cm 0.5cm 1.0cm},clip]{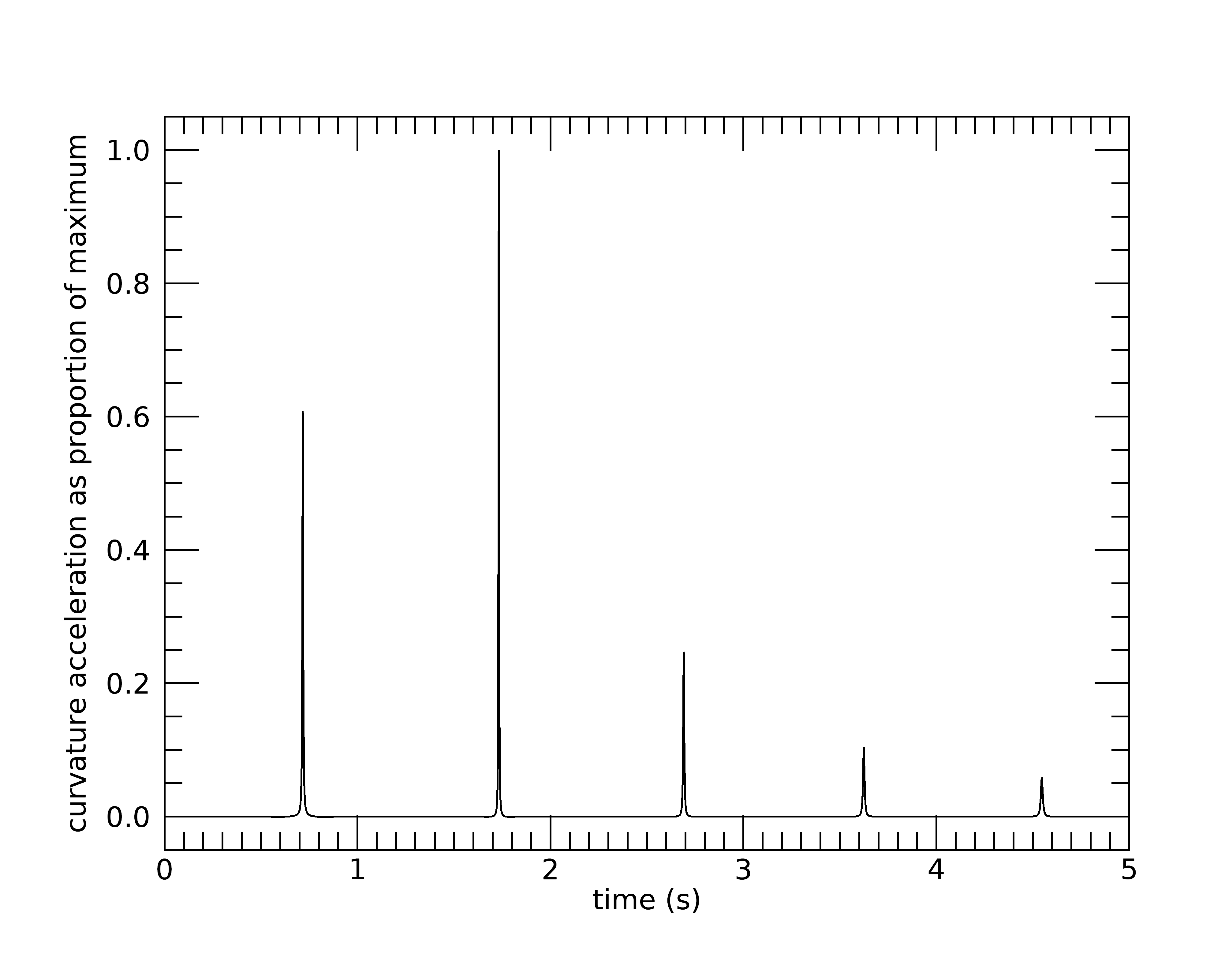}
    \includegraphics[width=0.99\columnwidth, trim={0.7cm 0.5cm 0.5cm 1.0cm},clip]{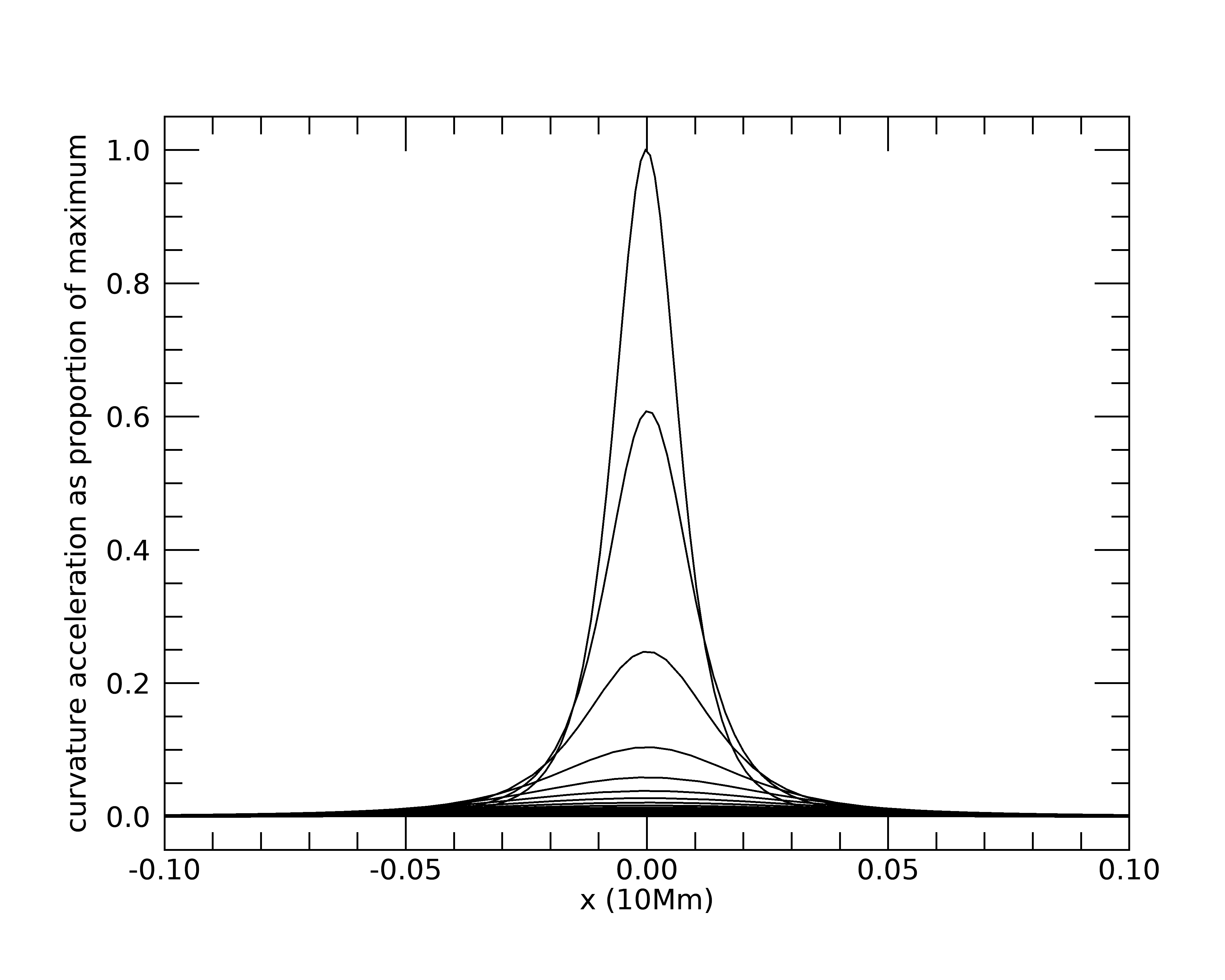}
    \caption{Fermi acceleration due to field line curvature (as a fraction of the maximum rate of Fermi acceleration) vs time [top] 
    and $x$ position [bottom] 
    for a particle orbit in 2D, started at (-0.4,1.25). We display the size of this term for only the first 5s of the 100s orbit and only over the range x = (-0.1,0.1) since the term stays close to zero outside of these ranges.}
    \label{fig:e5plots}
\end{figure}

Particle orbits which at some point pass through the minimum see far greater energy gains due to Fermi acceleration than those which do not. This shows a strong association between Fermi acceleration and fast collapse of the field lines at loop tops. Particle energisation at the loop top has previously been discussed in
\citet{giuliani:etal05}, \citet{grady:etal12}, and
\citet{eradat_oskoui:etal14},
and is supported by our results here. Inspection of the right hand side of Equation (\ref{eqn:dgammadt}) reveals that the key term for energisation by Fermi acceleration in the 
$\dot{\mathbf{R}}\cdot\mathbf{E}$ term for our model is

\begin{equation}
 \dot{\gamma}_{curv} =  \left(\frac{\textbf{b}}{B^{*}} \times u_{\parallel}\frac{d\textbf{b}}{dt}\right) \cdot \textbf{E},
    \label{eqn:curvterm}
\end{equation}

which is related to the curvature of the field lines.
%
%
This term causes
rapid acceleration over a very short period of time, specifically when an orbit is crossing the loop top. The contribution of this energisation term, relative to its maximum contribution in the particle orbit (starting at $(-0.4,1.25,0.0)$ in the 2D model), is shown over time and in space in Figure \ref{fig:e5plots}. The top panel shows how this term is only significant over very short periods of time and how it diminishes quickly as the speed of the field line collapse slows. The bottom panel shows how this term is strongest at the loop top and how it drops off quickly away from this region. For particles on orbits that start on field lines which are 
already
collapsed below the region of minimum magnetic field strength, the size of the spikes in energy gain related to this term are greatly diminished.  

Our dataset does not inform upon the energisation of wider particle populations, but does provide evidence that a large number of 
particle orbits 
(namely those 
starting
on more stretched field lines) gain significant energy due to Fermi acceleration. From the definition 
(\ref{eqn:curvterm}) of the term mainly responsible for Fermi acceleration, particles with a larger parallel velocity at the loop top obtain larger energy gains from curvature terms. Consequently, particles with smaller pitch angles will gain more energy this way compared to particles starting in the same position with a larger pitch angle. This indicates a possible association between particles that escape the CMT early and large energy gains resulting from Fermi acceleration at loop tops.  

Particle energy gains shown in Figure~\ref{fig:2Dinx} are relatively modest; the highest energy particles reach up to 2.5 times the initial energy. All particles orbits tested experience at least a 50\% increase in their energy, but only a minority see a final energy of more than double the initial energy. The Fermi effect seems to only lead to moderate particle acceleration, whilst the betatron effect has the potential to cause much stronger acceleration for a small subset of orbits. Such orbits are discussed in 
\citet{grady:etal12}. 
Orbits starting at the centre of the CMT ($x=0$) with a high pitch angle can reach energies of up to 40 times their initial energy. These orbits are not included in our simulations, since we want to investigate the energisation processes affecting the majority of orbits, not just those affecting quite particular orbits.  

In order to verify the association between Fermi acceleration and orbits starting on the most stretched field lines, we calculated a different set of particle orbits using the same 2D model. These orbits originate at the loop tops of field lines, initially stretched to varying degrees. All 121 orbits start at $x=0$ and on evenly spaced initial $y$ values between $1.5$ and $4.5$, i.e. a vertical straight line of positions. Initial energies remained at $5.5$keV, but, in order to better observe Fermi acceleration, initial pitch angles were reduced to $20^{\circ}$, with this initial pitch angle being high enough to keep orbits trapped and low enough to prevent all orbits from being constrained to a narrow region of the field.  

\begin{figure}
    \centering
    \includegraphics[width=\columnwidth,trim={0.8cm 0.7cm 1.0cm 1.2cm},clip]{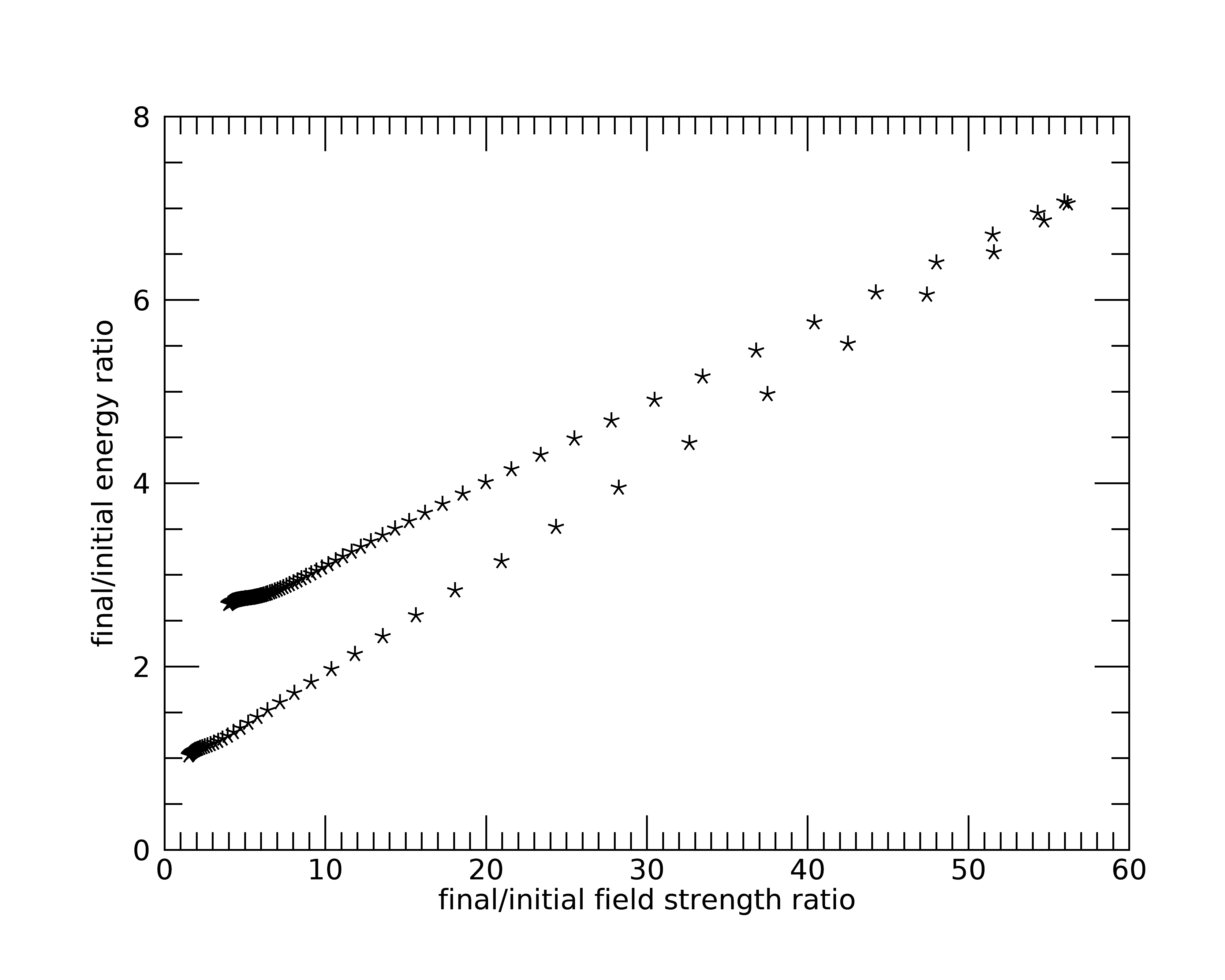}
    \caption{Ratio of final to initial energy against final to initial field strength for orbits with different initial $y$ values in the 2D model. Initial energies were set at $5.5$keV and initial pitch angles at $20^{\circ}$.}
    \label{fig:2Diny}
\end{figure}

The results are presented in Figure \ref{fig:2Diny} and show that for these orbits, initial conditions lying on field lines above the minimum in the field strength are once again 
associated with Fermi acceleration. The chart shows a curve with two branches which meet at a point corresponding to the orbit starting closest to the 
minimum in the magnetic field 
%
strength.
The upper branch corresponds to trajectories starting above the minimum in the field strength and the lower curve corresponds to trajectories starting below it. By inspecting points representing orbits on the two branches in Figure \ref{fig:2Diny} with similar field strength ratios (i.e. vertically aligned points), we can directly compare orbits with almost identical expected energy gains due to betatron acceleration. One notices a gap in particle energies, which increases from right to left, which is due to Fermi acceleration energising particles starting on more and more stretched field lines. 

\subsubsection{Speeding up CMT collapse}

With the impact of Fermi acceleration in our model established, we sought to better understand how the speed of the CMT field collapse impacts energy gains. The bigger spikes in energy gains associated with weaker regions of the field suggest a CMT configuration in which field lines collapse faster may see a greater contribution from Fermi acceleration towards energies for particles starting on field lines above the minimum in the field strength. 

To investigate the effect, we consider the stretching term detailed in Equation (\ref{eqn:2dyinf}). 
The only time-dependence in this transformation comes from the $(at)^b$ terms, where the current model 
has $a=0.4$ and $b=1.0$. The collapse of field lines can be 
sped up by increasing the value of $b$ and 
the initial state of the field can be preserved between 
models by adjusting $a$ to ensure that $(at)^{b}$ 
takes the same value at $t = 1.05$ (the initial normalised time). 
To investigate this we double $b$ ($b = 2.0$) which implies 
%
%
$a=\sqrt{0.4/1.05}$. 
In this faster setup, field lines move further in the same amount of time, so a stronger field is produced, hence particles will gain more energy from betatron acceleration.
Since we are only interested in the differences in Fermi acceleration 
resulting from a faster collapse of field lines, we simulate 
 up to different 
times for each setup, so that they both relax to the same field 
configuration by the end of each simulation. 
To reach the same field configuration, $(at)^{b}$ must be identical at 
the final times of both simulations. This can be achieved by 
running the faster simulation for $100$s and running 
the slower simulation, studied in section \ref{ssec:2Dresults}, for $295$s. 






\begin{figure}
    \centering
    \includegraphics[width=\columnwidth,trim={0.8cm 0.8cm 1.32cm 1.55cm},clip]{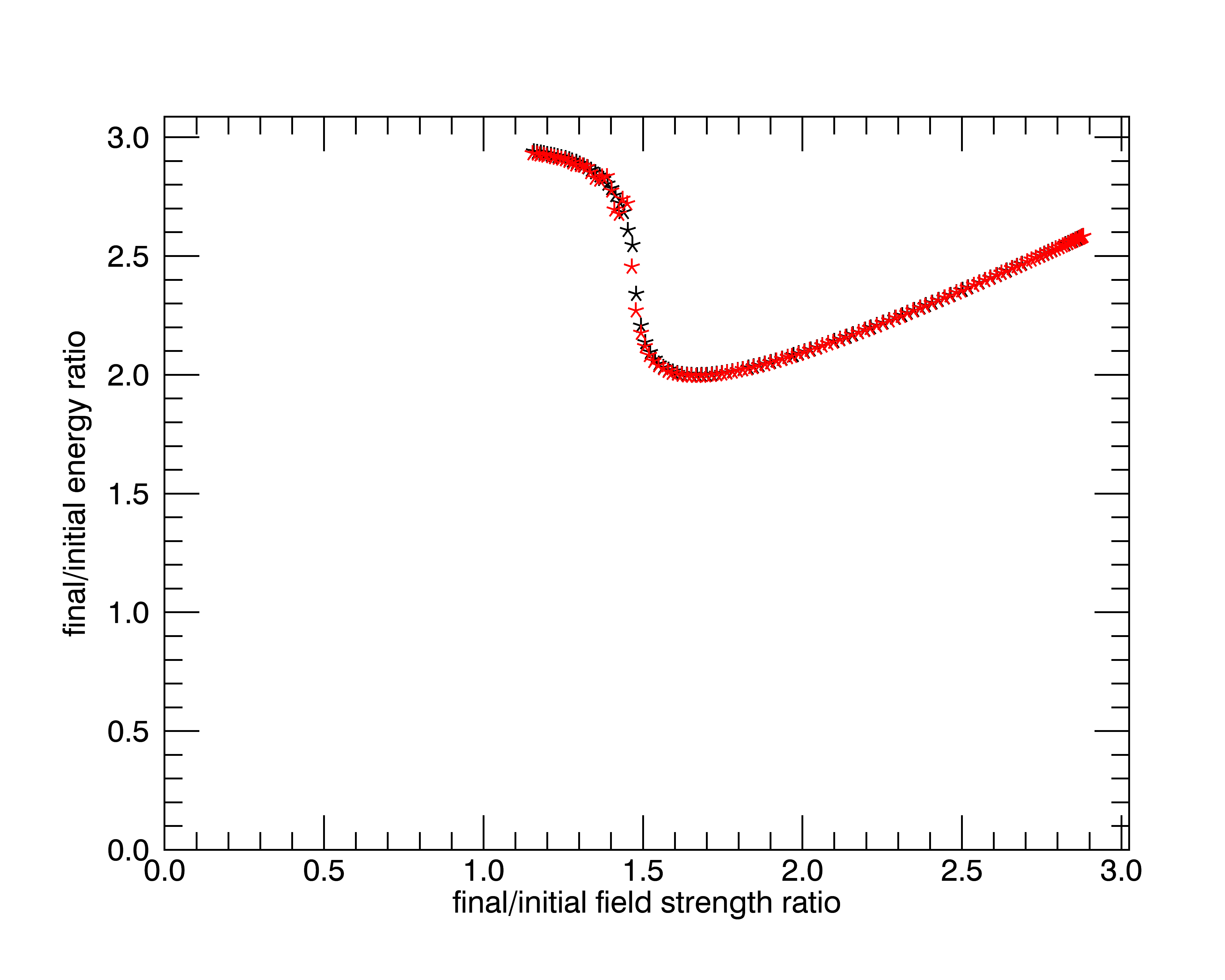}
    \caption{Comparison of results for the initial conditions of the standard case used in the 2D CMT model with the regular configuration run over $295.2$s (black) with the result of the faster collapse CMT model run over $100$s (red).}
    \label{fig:2dspeedcomp}
\end{figure}

Figure \ref{fig:2dspeedcomp} shows the scatter plot generated from the above simulations. 
For each simulation the data points yield the same curve, 
indicating that a faster collapse of field lines will not lead to a bigger overall energy 
contribution from Fermi acceleration. Rather, the energy gained due to Fermi acceleration seems to be more closely related to the overall distance that field lines collapse through. The minimum in the field strength is associated with the `transitional region' (still located between 0.9 and 1.0 in the field strength ratio) because, for initial positions evenly spaced in $x$ with constant $y$, field lines will be significantly more stretched in the $y$ direction if they lie above the minimum. 

Another interesting feature of Figure \ref{fig:2dspeedcomp} is in the `transitional region' itself. 
Orbits represented in this region are the only ones where clear differences in energies are visible between particles starting at the same positions (vertically aligned points) in the faster and slower setup.
Since these orbits started on field lines close to the minimum in the 
%
%
magnetic
field
strength, a 
possible explanation for the energy discrepancy is that some particles may gain 
less energy than others by not passing through the loop top at a time when it 
is collapsing fastest. This opens up the possibility of particles missing 
out on possible energisation by spending time in the loop legs whilst the field 
line is collapsing at its fastest rate. Most particles 
%
%
%
make loop top passes often enough for this to have a minimal effect on 
particle energies in this family of models, but it may be of more importance for 
%
%
%
CMT models
with different properties. 
We do not see this effect for particle
orbits 
displayed
in the top left region, 
because the field lines on which these orbits start do not pass the minimum early in the simulation, so do not collapse as fast. 
As field lines collapse, the 
minimum of the field strength becomes larger in value, corresponding to slower field line collapse. 
Particles starting on such field lines will almost certainly not be 
caught in the loop legs at a time when the loop top is collapsing 
fast enough to affect particle energies significantly. 

\subsection{2.5D case with guide field}
\label{ssec:2_5Dresults}


Using MHD simulations based on a particular initial magnetic field configuration \citet{birn:etal17} showed the importance that a guide field component can have on the possible energy gains by the betatron mechanism. 
In order to connect this with the energisation processes seen in CMTs it is vital to extend our models to 2.5D and 3D. 
Even in 2D CMT models, orbit calculations have to be fully 3D because trajectories have drift components in the invariant direction,
although
the observed motion in the third direction is minimal (in the order of a few kilometres in our standard normalisation) and mostly cancels between loop top passes.
First extending
our models to 2.5D 
%
%
%
allows us to add a guide field component, so that we have more realistic models with an added degree of complexity. 

In a 2.5D we include a time- and space-dependent guide field component, $B_{z}$. As mentioned in Section \ref{ssec:CMTtheory}, this component is given as:
\begin{equation}
    B_{z}=\frac{\partial y_{\infty}}{\partial y} B_{z final}.
\end{equation}
The term $B_{z final}$ is a constant in our model. The $B_{z}$ component tends towards $B_{z final}$ from below as the field collapses towards the potential field state. The $B_{x}$ and $B_{y}$ components and $A_{\infty}$ are unchanged from the 2D case, allowing for easier comparison between the 2D and 2.5D models (since the 2.5D model will reduce to the 2D model if $B_{z final} = 0$). To see the impact of the $B_{z}$ component, or guide field, on particle energisation, we will compare scatter plots like those shown in the 2D case, as well as 2.5D cases with $B_{z final}$ set to $0.005$T and $0.01$T. These values have been chosen to highlight differences in particle energisation. For all simulations we again initialise 121 orbits with initial energies of $5.5$keV and pitch angles of $60^{\circ}$ on points equally spaced between $x=-0.5$ and $x=0$ and with $y = 1.25$ (a horizontal line of points). 

\begin{figure}
    \centering
    \includegraphics[width=\columnwidth,trim={0.8cm 0.8cm 1.32cm 1.55cm},clip]{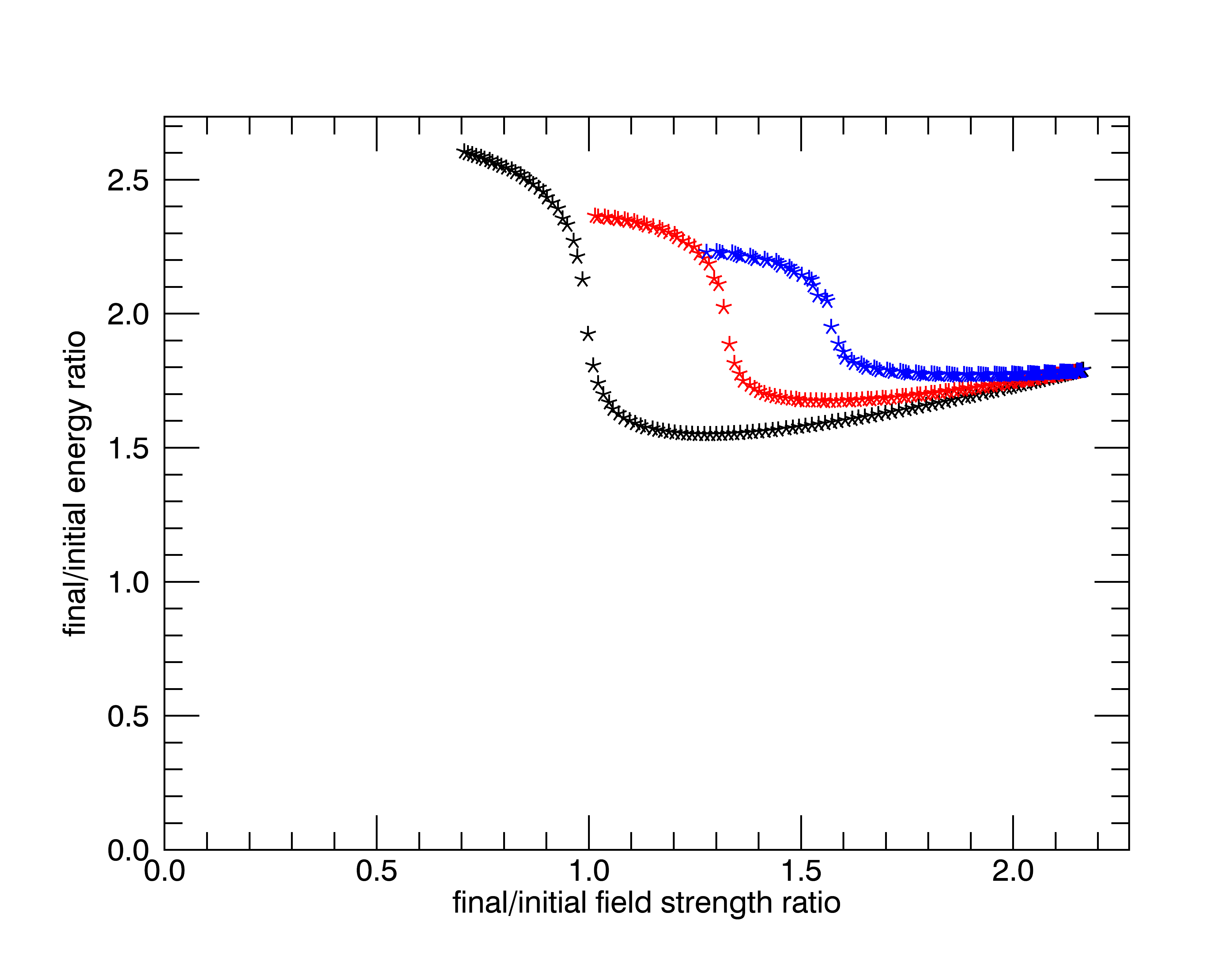}
    \caption{Ratios of final to initial energy against final to initial field strength for the 2D case (black) and the 2.5D cases with $B_{z final}=0.005$T (red) and $B_{z final}=0.01$T (blue). Initial conditions are as described in Section \ref{ssec:2_5Dresults}.}
    \label{fig:bzcompinx}
\end{figure}

In Figure \ref{fig:bzcompinx} we present the ratios of final to initial energies against the ratios of final to initial field strength for these conditions. The 
%
%
case
with $B_{zfinal} = 0$ 
(i.e. the 
%
%
2D
case already displayed in Figure \ref{fig:2Dinx}) is presented in black, the
%
%
case
with $B_{zfinal} = 0.005$T is in red and the 
%
%
case
with $B_{zfinal} = 0.01$T is in blue. 
%
%
The general structure of our data points is preserved for varying $B_{z final}$, but clear differences have emerged when a guide field is introduced and varied. Particles in 
%
cases with a stronger guide field are gaining more energy from betatron acceleration and less from Fermi acceleration. The introduction of a $B_{z}$ component causes particle orbits to both start and end the simulation in a stronger field than orbits with the same starting points in the 2D model. The expected increase in betatron acceleration due to a stronger final field will be offset by a decrease in the magnetic moment due to the stronger initial field. 
Particles can be expected to gain more energy due to betatron acceleration when a guide field is introduced as long as the final to initial field strength ratio is increased with its introduction, giving the condition:

\begin{equation}
  \frac{(B_{x,end}^{2}+B_{y, end}^{2}+B_{z,end}^2)^{1/2}}{(B_{x,init}^{2}+B_{y,init}^{2},B_{z,init}^2)^{1/2}} > \frac{(B_{x,end}^{2}+B_{y, end}^{2})^{1/2}}{(B_{x,init}^{2}+B_{y,init}^{2})^{1/2}}, \nonumber
\end{equation}
where ($B_{x,init}$,$B_{y,init}$,$B_{z,init}$) is the magnetic field at the initial position of the particle and ($B_{x,end}$,$B_{y,end}$,$B_{z,end}$) is the field on the final loop top pass. Note that $B_{z final}$ and $B_{z,end}$ do not necessarily take the same value as the field may not have totally collapsed by the end of the simulation. The above expression will be satisfied if:

\begin{equation}
   \frac{B_{z,end}}{B_{z,init}} > \frac{(B_{x,end}^{2}+B_{y,end}^{2})^{1/2}}{(B_{x,init}^{2}+B_{y, init}^{2})^{1/2}}.
   \label{eqn:betatroncondition}
\end{equation}
Here the initial $y$ position is constant, so $B_{z init}$ will be identical for all the particle trajectories since $\frac{\partial y_{\infty}}{\partial y}$ is only dependent on $y$ and $t$. Whilst not identical, the final field components will be similar for all orbits, because, within the $100$s simulation time, these field lines will all have mostly collapsed. The $(B_{x,init}^{2}+B_{y,init}^{2})^{1/2}$ term however, will vary significantly between orbits. Orbits that start further out in $x$ are located further into the loop legs. As a result, they will start in a stronger region of field than those closer to the centre of the CMT. Consequently, Equation (\ref{eqn:betatroncondition}) will be more easily satisfied for particles starting further out in $x$, and therefore starting on the most stretched field lines. This leads to greater energy gains due to betatron acceleration for particles on orbits starting on more stretched field lines, as shown in Figure \ref{fig:bzcompinx}. The curves begin further to the right for larger $B_{z,final}$, as the ratio of field strengths is larger, also as a result of Equation (\ref{eqn:betatroncondition}) being satisfied.  

We also observe from Figure \ref{fig:bzcompinx} that a stronger guide field reduces the energy gain due to Fermi acceleration for particles on more stretched field lines. Even though points are shifted to the right, indicating stronger betatron acceleration, many particles actually gain less overall energy, meaning that the Fermi contribution is reduced. 
We remark that the reduction of the contribution of Fermi acceleration with a stronger guide field is reminiscent of results based on particle-in-cell simulations by \citet{dahlin:etal14}, albeit for systems with different scales.

We have established that the main contribution to Fermi acceleration comes from the curvature of field lines and this explains the reduction of this type of energisation for larger values of $B_{z,final}$. Since the coordinate transformations $x_{\infty}$ and $y_{\infty}$ are identical for the 2D and 2.5D models, both models will give the same magnetic field projection onto the $x-y$ plane. With the introduction of a $B_{z}$ component, loop tops are flattened out
%
%
in the $z$-direction. 
This reduction in curvature reduces the energising effect of Fermi acceleration.  

\begin{figure}
    \centering
    \includegraphics[width=\columnwidth,trim={0.8cm 0.7cm 1.0cm 1.2cm},clip]{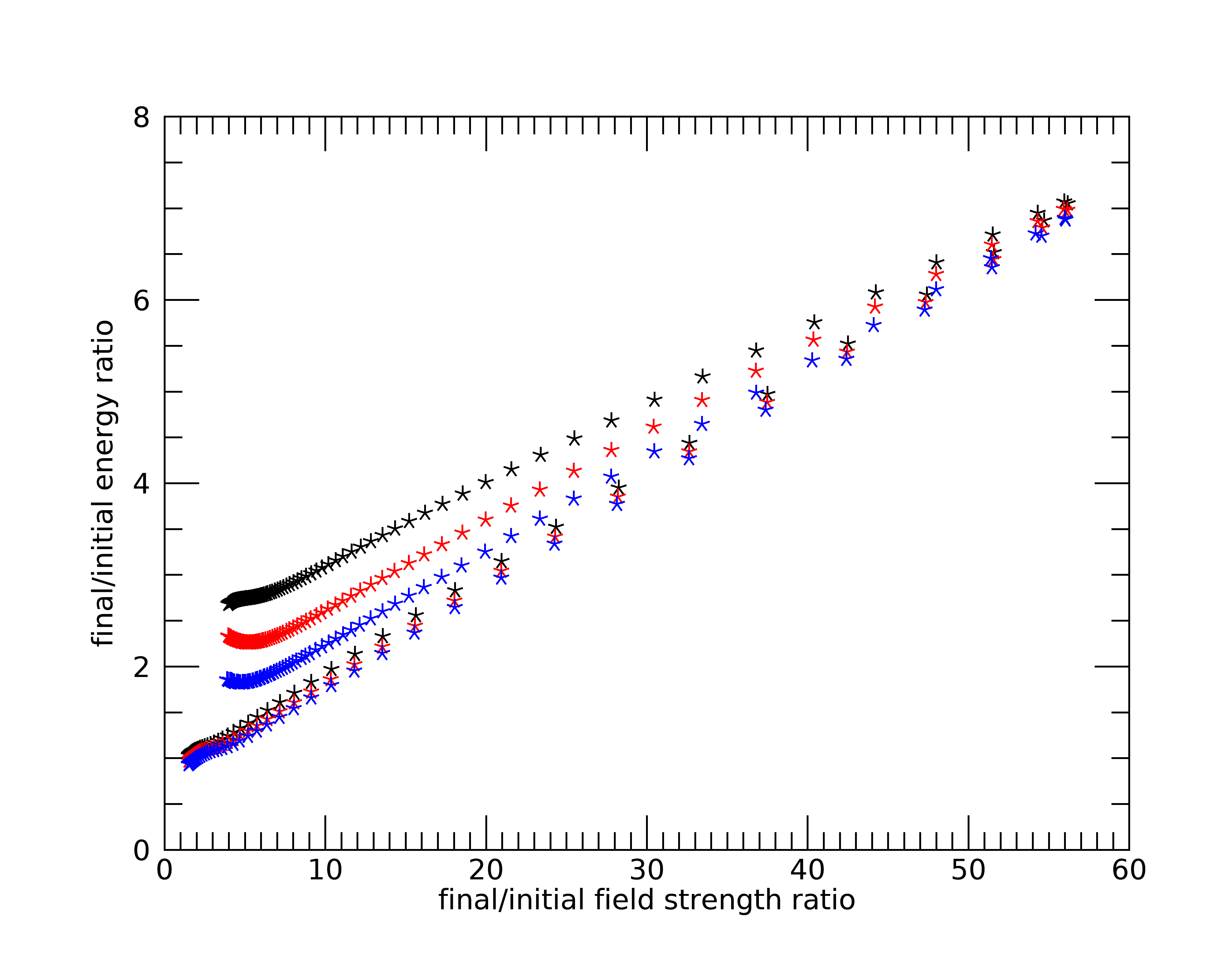}
    \caption{Results
    from the 2D model (black) and the 2.5D model with $B_{z final}=0.005$T (red) and $B_{z final}=0.01$T (blue). All data points have initial positions of $x = 0$ with $y$ varying uniformly between $1.5$ and $4.5$. Initial energies are 5500 eV and pitch angles are $20^{\circ}$.}
    \label{fig:bzcompiny}
\end{figure}

To investigate this further we retain the same field configurations ($B_{z,final} = 0$T, 0.005T, 0.01T) and initialise orbits with a $5.5$keV energy, $20^{\circ}$ pitch angle and initial positions of $x = 0$ and $y$ evenly spaced over 121 orbits between $1.5$ and $4.5$ (the same vertical line of 
%
%
initial positions
used to generate the data displayed in Figure \ref{fig:2Diny}). 
The results are shown in Figure~\ref{fig:bzcompiny} with existing data from the 2D model shown in black, data with $B_{z final}=0.005$T in red and data with $B_{z final}=0.01$T in blue. For these orbits, $B_{z}$ evolves in the same way as the other magnetic field components and hence the introduction of the guide field does not lead to any increase in betatron acceleration. We do however, observe a reduction in energy gains due to Fermi acceleration for stronger guide fields. There is a clear gap in the three upper branches in the figure, representing those orbits that start on field lines stretched above the minimum in the field strength. This gap represents a reduction in energisation due to Fermi acceleration for stronger guide fields. Again, this results from the flattening of loop tops by the guide field. Interestingly, a smaller, but still visible gap in energy gains can be seen in the three lower branches. This suggests that particles with orbits that start on field lines lying below the minimum in the field are still being energised due to Fermi acceleration, just less so than those starting on more stretched field lines. In particular, orbits in the 2D model represented on this lower branch, which see field strength ratios between 10 and 20, have at least 10\% of their particle energisation 
%
%
caused
by Fermi acceleration. 

\subsection{3D case}
\label{sec:3Dresults}

\begin{figure}
    \centering
    \includegraphics[width=\columnwidth, trim={0 1cm 0 4cm},clip]{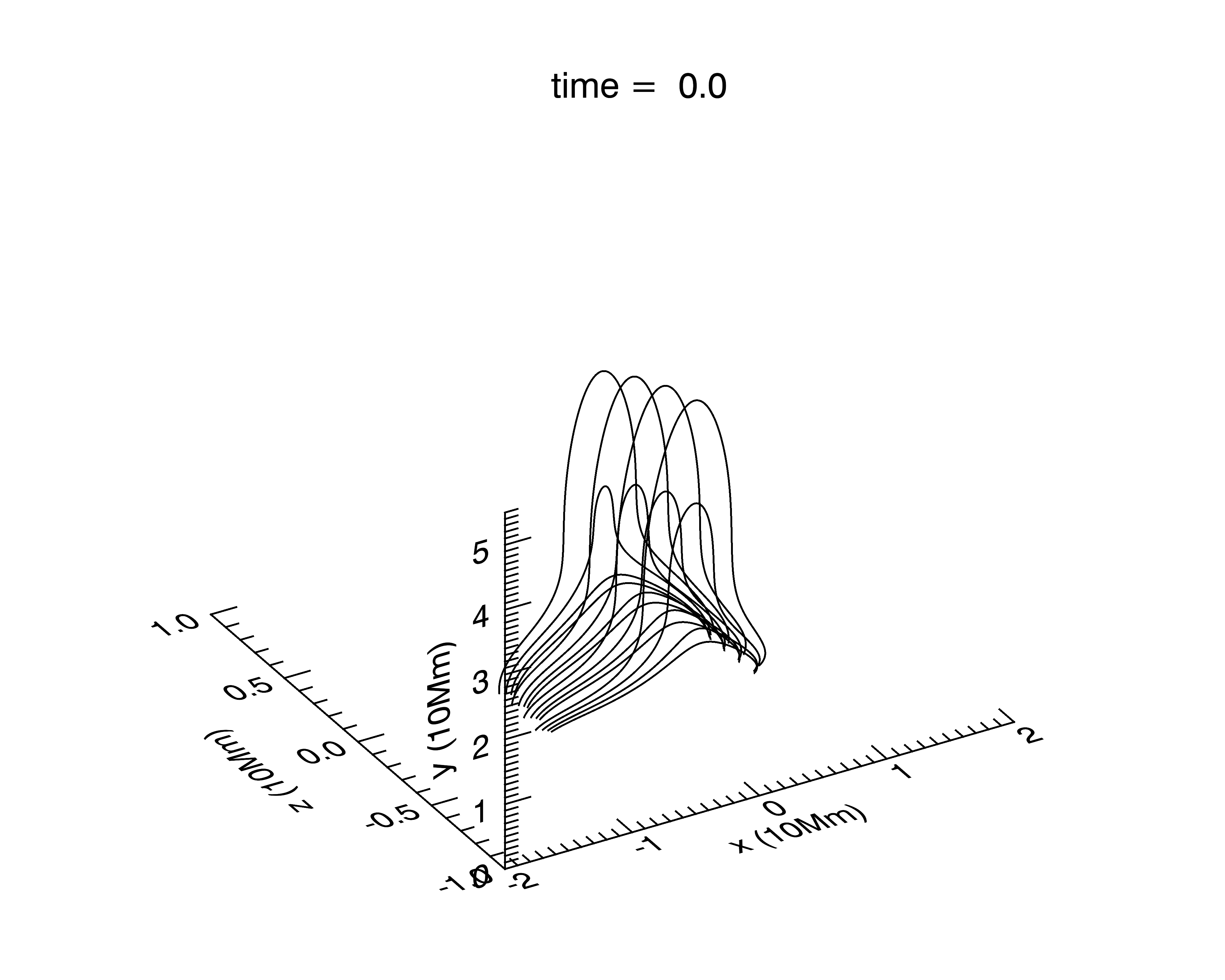}
    \includegraphics[width=\columnwidth, trim={0 1cm 0 4cm},clip]{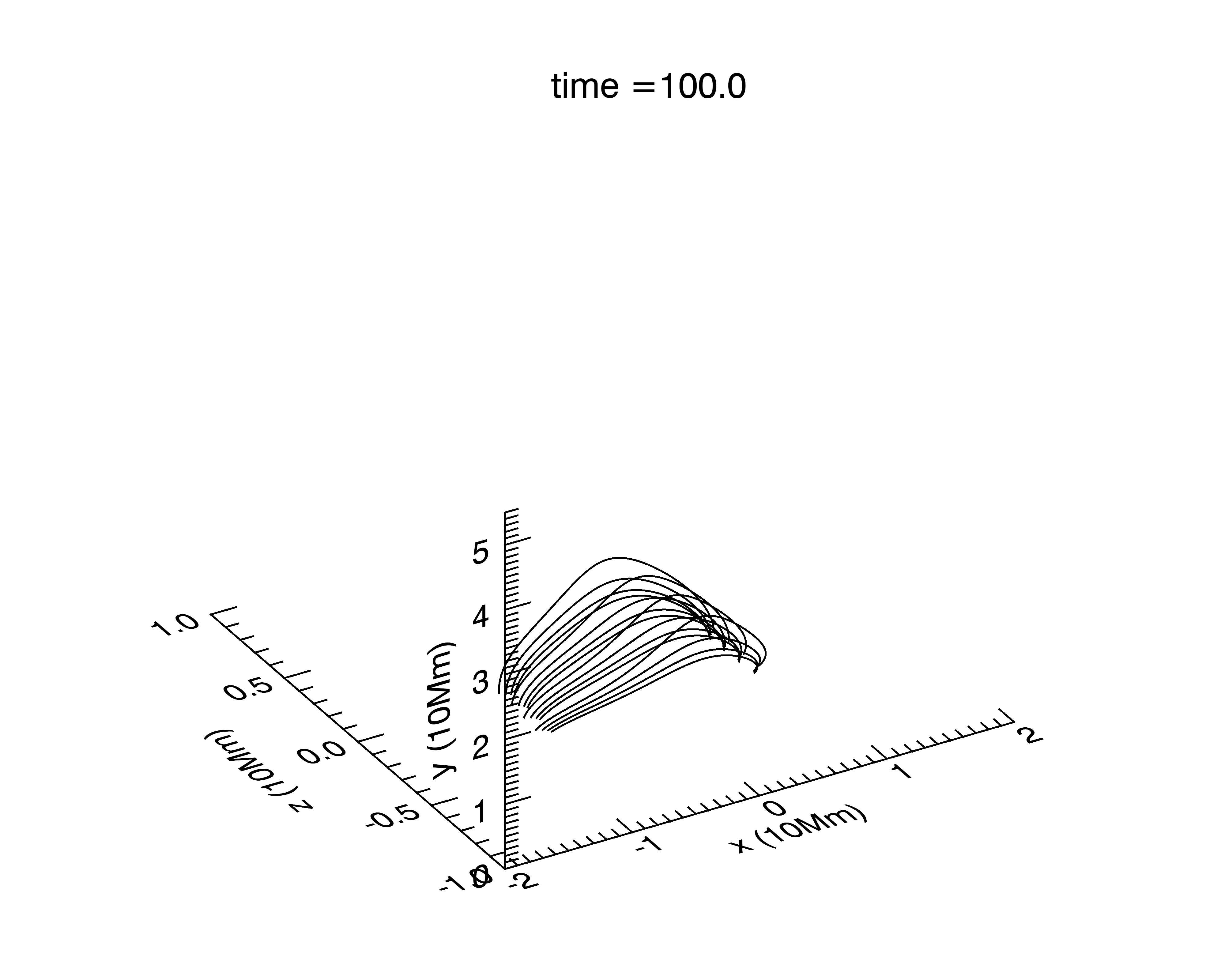}
    \caption{Field lines plots for the 3D model at $t=0$s (top) and $t=100$s (bottom), with field lines traced from the initial positions used for particle orbits.}
    \label{fig:3dflines}
\end{figure}

We will now present results from our
%
%
modified
3D model as detailed in Section \ref{sec:basic}. 
Although extending our CMT models to 2.5D allows us to include a guide field,
extending our model to 3D presents the opportunity to
make the CMT magnetic fields even more realistic. In particular, one key difference between 2D/2.5D and 3D models is the dependence of the magnetic field strength on
the distance from the source regions, which drops off faster in 3D than in 2D and 2.5D.
One question arising in this context is how this might affect particle orbits and related
changes in energy.
Another key difference between the 3D model and those of lower dimensions, is that the 3D field is no longer invariant in the $z$ direction, making the initial $z$ value of a particle orbit a consequential initial condition. Additionally, higher order drift terms could potentially move orbits onto field lines with different properties. 

As our standard case we take $\delta=1.0$, $a_{y}=1.0$ and $y_{h}=0.0$, in order to generate a twist in the field that reduces with increasing $y$. We again consider a range of initial positions that will see orbits start on a set of field lines with varying degrees of initial stretching and twisting. This can be achieved by taking initial positions on a square grid in the $x-z$ plane, with $y=1.25$ and with $x$ and $z$ both varying between $-0.5$ and $0$ with 11 evenly spaced steps. Some of the field lines that the orbits initially start on are shown in the top panel of Figure \ref{fig:3dflines}. In 3D, our standard case uses initial energies of $5.5$keV and initial pitch angles of $70^{\circ}$. This is increased from the lower dimensional cases to ensure that all of the orbits remain trapped for the duration of the simulation. This will avoid confusion that may arise from recording final energies and field strengths at very different times between orbits. 

\begin{figure}
    \centering
    \includegraphics[width=\columnwidth,trim={0.8cm 0.8cm 1.32cm 1.55cm},clip]{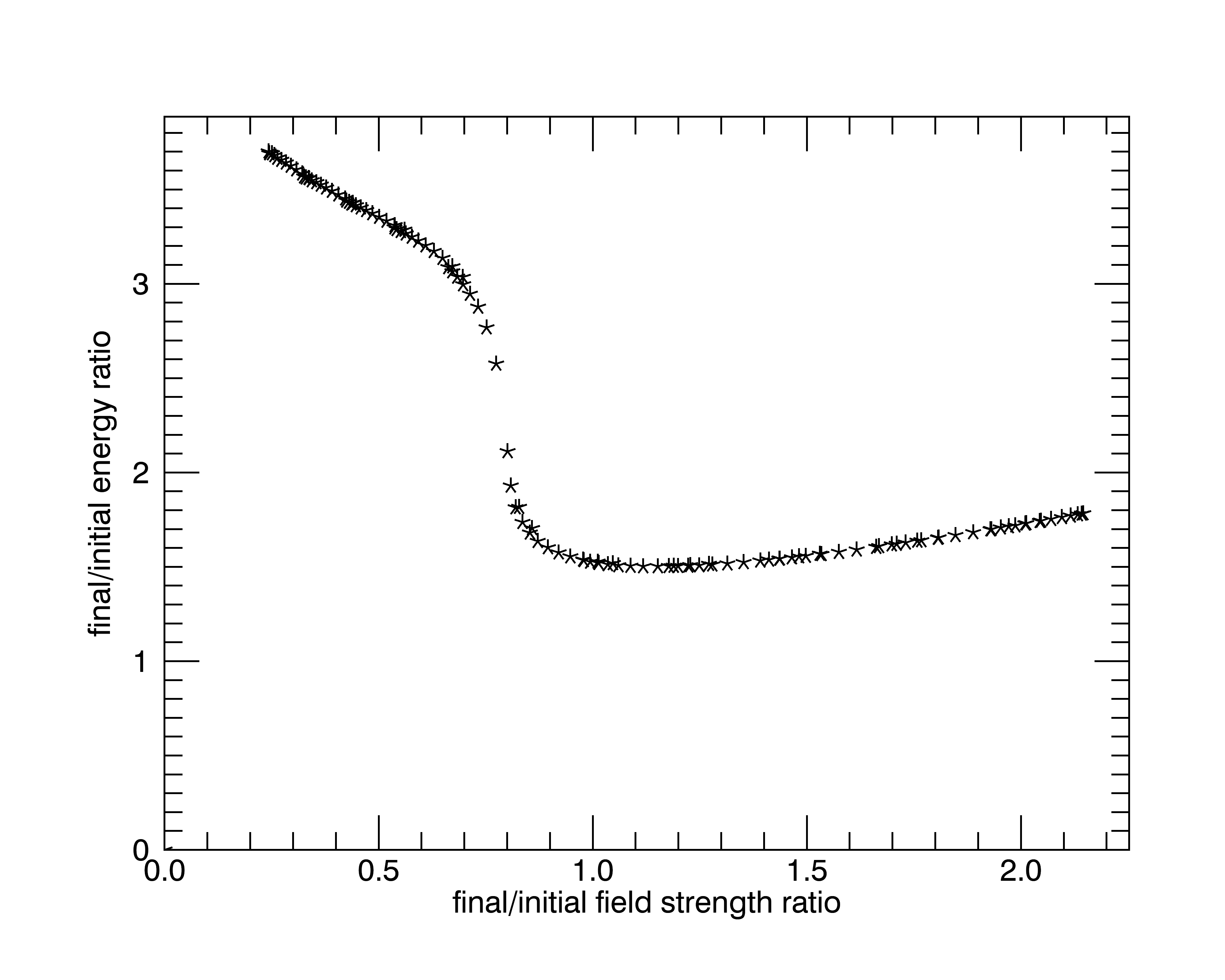}
    \caption{Ratio of final to initial energies against final to initial magnetic field strengths for 121 orbits initialised on the square grid as described in the standard case for our 3D initial conditions.}
    \label{fig:3dscatter}
\end{figure}

Results for these initial conditions are presented in Figure \ref{fig:3dscatter}. 
The three distinct regions of behaviour identified in 2D/2.5D simulations are readily apparent in Figure \ref{fig:3dscatter}.
The section to the right where betatron acceleration dominates is located to the right of the point where the field strength ratio is 0.9; the top 
left section where particles see significant energy gains due to 
Fermi acceleration is found for a field strength ratio of less than 0.7; and a broader `transitional region' linking the two is found between field ratios of 0.7 and 0.9, where the energy ratio drops from 3.0 to 1.5. 
Similarly to previous cases, orbits which gain energy mainly through betatron acceleration 
initially lie on the most collapsed field lines, 
whilst orbits which gain the most energy due to Fermi 
acceleration start on the most stretched field lines. 

It is perhaps 
surprising to see the `transitional region' preserved so well for 
3D simulations, in spite of the more complicated nature of the minimum 
in the field strength for the 3D CMT model with a twisted field. In 
%
%
the
2D and 2.5D
%
%
models, 
the minimum of the field is located at a specific point in the $x-y$ plane. 
In 3D a minimum point in the field strength still exists, but contrary to 2D/2.5D,
the majority of field lines will never pass through this point. Instead, 
for our simulations, the relative minimum in the field strength relevant to a particle orbit is 
the weakest point in the region of the magnetic field swept out by its field line as it collapses. 
The clear `transitional region' in the 3D model may be a result of 
the invariance of the coordinate transformation responsible for 
stretching, $y_{\infty}$, in both $x$ and $z$. We have here opted for keeping
the transformation relatively simple and as close as possible to the lower dimensional cases,
but this aspect of the 3D CMT model could be improved upon in future work.

Figures \ref{fig:2Dinx} and \ref{fig:3dscatter} show 
that the particles in the 3D simulation have gained more energy 
than those in the 2D model, though direct comparisons of test particle orbits 
between the models are difficult since there is no clear way to compare 
initial particle positions and field lines. 
Far more particle orbits would need to be considered for each model to 
compare energisation in the 2D and 3D CMT models. 
However, for the number of orbits used here, 
%
%
interesting details are visible when comparing Figures \ref{fig:2Dinx} and \ref{fig:3dscatter}. 
In particular, particles on orbits that begin on the most stretched 
field lines in the 3D model gain more energy due to Fermi acceleration than 
their 2D counterparts. This is surprising, since the 3D particle orbit 
simulations utilised higher initial pitch angles ($70^{\circ}$) compared to the 2D case ($60^{\circ}$). 
Increasing the initial pitch angle 
reduces $u_{\parallel}$ across the orbit, 
including at the looptop. From Eq (\ref{eqn:curvterm})  
this should reduce the energy gained due to Fermi acceleration, 
so it seems that the 3D model can more efficiently energise particles this way. 

\begin{figure}
    \centering
    \includegraphics[width=\columnwidth,trim={0.8cm 0.8cm 1.32cm 1.55cm},clip]{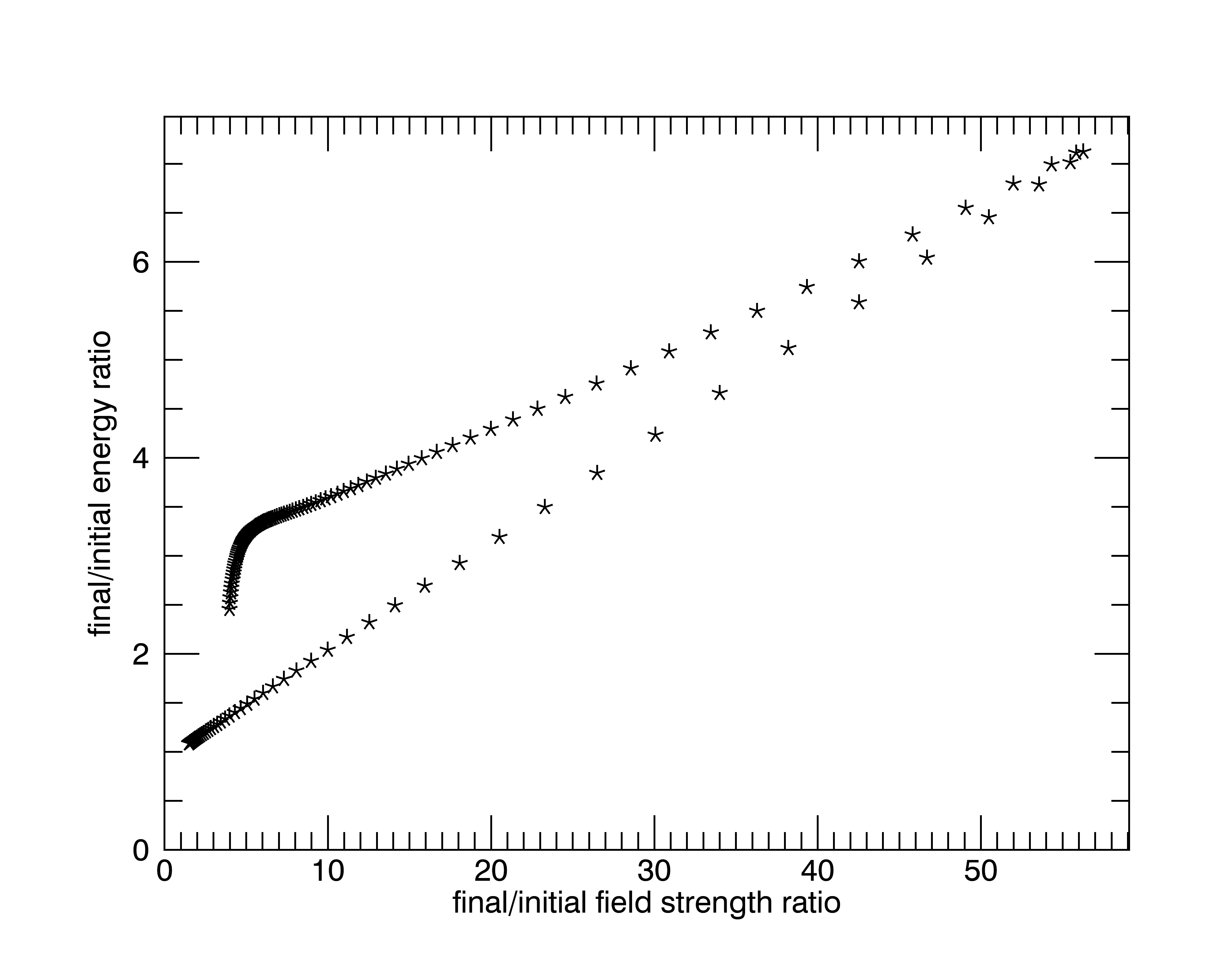}
    \caption{Ratio of energy gain vs ratio of field strength increase for orbits starting at $x = 0$, $z = 0$ and $y$ varying between $1.0$ and $4.5$. The initial pitch angle is $20^{\circ}$.}
    \label{fig:3diny}
\end{figure}

For our 3D model, we also present the case where orbits start at the centre of the CMT with 
different heights. As our initial conditions, we set initial $x$ and $z$ values of $0$ 
and vary the initial $y$ value between $1.0$ and $4.5$, calculating 121 equally spaced orbits. 
The initial pitch angle is chosen as $20^{\circ}$ so that we can better compare results 
with Figure \ref{fig:2Diny}. These initial positions allow us to investigate the energies of particles starting in the weakest regions of the magnetic field and the low pitch angle prevents some orbits from remaining very close to the loop top (where the field is weakest). 

These results are presented in Figure \ref{fig:3diny} where we see a similar curve to that of Figure \ref{fig:2Diny}. 
Again, two straight lines meet at the point of maximum energy gain. Individual inspection of particles shows that 
the lower line corresponds to orbits that start below the minimum in the field 
strength along $x,z = 0$, and the upper line corresponds to orbits that start above 
the minimum. We conclude that the extra energy gains result from Fermi acceleration due to 
loop top crossings, especially when the loop top is near the minimum in the field strength. 
The orbits represented 
in the upper line show energy gains that 
cannot be
%
%
explained by betatron acceleration alone. 

The other key feature of Figure \ref{fig:3diny} is the tailing off of energies for orbits 
presented on the left side of the upper line. These represent the orbits starting highest 
in the CMT. Closer inspection shows that these particles show lower energy gains than 
expected because the field lines that they lie on do not collapse as far as those of the 
other orbits presented in this figure. Because of this particles gain less energy due to betatron and Fermi acceleration as
%
%
the loop top has not moved as far
during the field line collapse.

\subsubsection{Different initial energies}
\label{ssec:3denergies}

So far, we have 
%
%
only
investigated particles with initial energies of $5.5$keV. 
%
%
%
We expect that particles will be injected into the CMT with a wide range of energies. For example, particle populations that have been pre-heated or pre-accelerated in the
reconnection region (which is not included in our model), could have 
significantly higher initial energies. On the other hand, since the 
reconnection process changes magnetic connectivity not just locally, but on a larger scale; 
particles could enter the CMT without having passed through the reconnection region and could have much lower initial energies. Hence, we would like to investigate different initial particle energies from the $5.5$keV used so far. To investigate how 
initial energies influence further energisation, we consider two additional
data sets which use the initial conditions used to generate Figure \ref{fig:3dscatter}, but have initial energies
of $55$keV and $0.55$keV, respectively, i.e. one order of magnitude difference from the
nominal initial energy in either direction.

\begin{figure}
    \centering
    \includegraphics[width=\columnwidth,trim={0.8cm 0.8cm 1.2cm 1.55cm},clip]{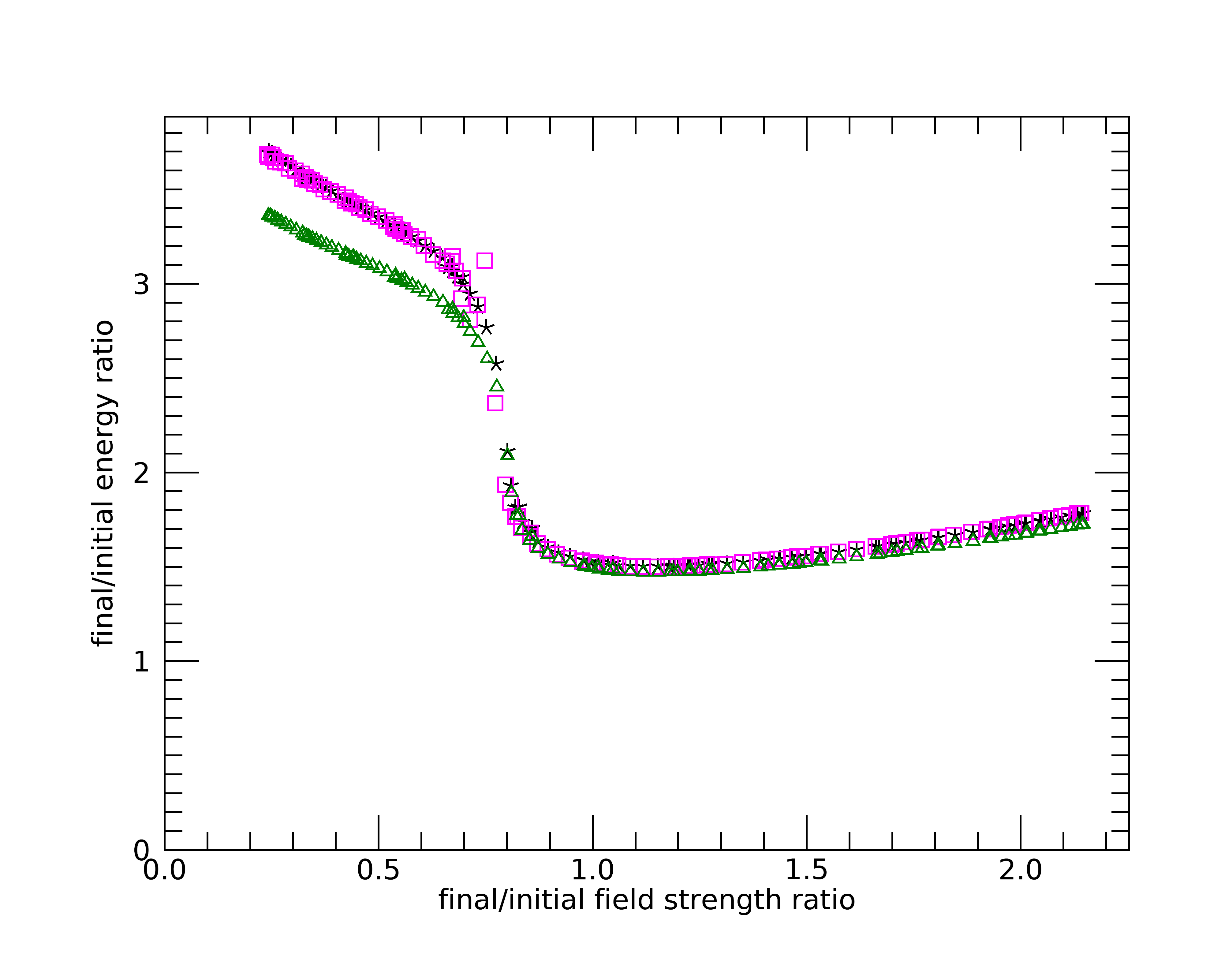}
    \caption{Energy ratios vs field strength ratios
    for initial energies of $0.55$keV (magenta 
    %
    %
    squares), 
    $5.5$keV (black
    %
    %
      asterisks) and $55$keV (green
      %
    %
      triangles) using the initial positions and pitch angles for the 3D model with initial conditions described in Section \ref{ssec:3denergies}.}
    \label{fig:3denergycomp}
\end{figure}

These results are presented in Figure \ref{fig:3denergycomp}. The general structure of the data points is preserved for each initial energy, with the data points for initial energies of $0.55$keV 
%
%
(magenta squares) and $5.5$keV 
%
%
(black asterisks) lying on almost identical curves. This indicates that, for non-relativistic particles, orbits which start in the same positions and with the same pitch angles will experience similar energy gains relative to their initial energy. 
This is to be expected,
%
%
because 
energy gains due to betatron acceleration depend on $\mu$ and therefore $v_{\perp init}$ and Fermi acceleration 
is dependent on $u_{\parallel}$. The only noticeable differences in these datasets emerge 
%
%
within
the `transitional region', where vertically aligned points (representing orbits with the same initial positions, can be separated by significant differences in their final energies. These suggest that, for orbits starting on field lines close to the minimum in the field, whether or not an orbit crosses the looptop at the time of fastest collapse significantly impacts the final energy. Aside from an increase in `noise' in this region, the shape of the `transitional region' is very similar for the two initial energies. 

The line of green 
%
%
triangles
in Figure \ref{fig:3denergycomp} represent the particles that start with energies of 55keV. These particles gain a lower factor of energy than their 
%
%
non-relativistic 
counterparts. The difference in the energy ratio is clearest in the top left of the Figure, where the most energetic particles in each data set are represented. 
These results support the findings of 
\citet{eradat_oskoui:neukirch14},
who
found quantitative differences in energy gains when comparing relativistic and non-relativistic regimes. These moderately relativistic particles are still  
substantially
%
%
energised, with energies of orbits starting on the most stretched field lines increasing by a factor of between 2.8 and 3.5.

\subsubsection{Effect of 3D model parameters}
\label{ssec:3Dparameters}

\begin{figure}
    \centering
    \includegraphics[width=\columnwidth, trim={1.5cm 1.5cm 1.5cm 1.5cm},clip]{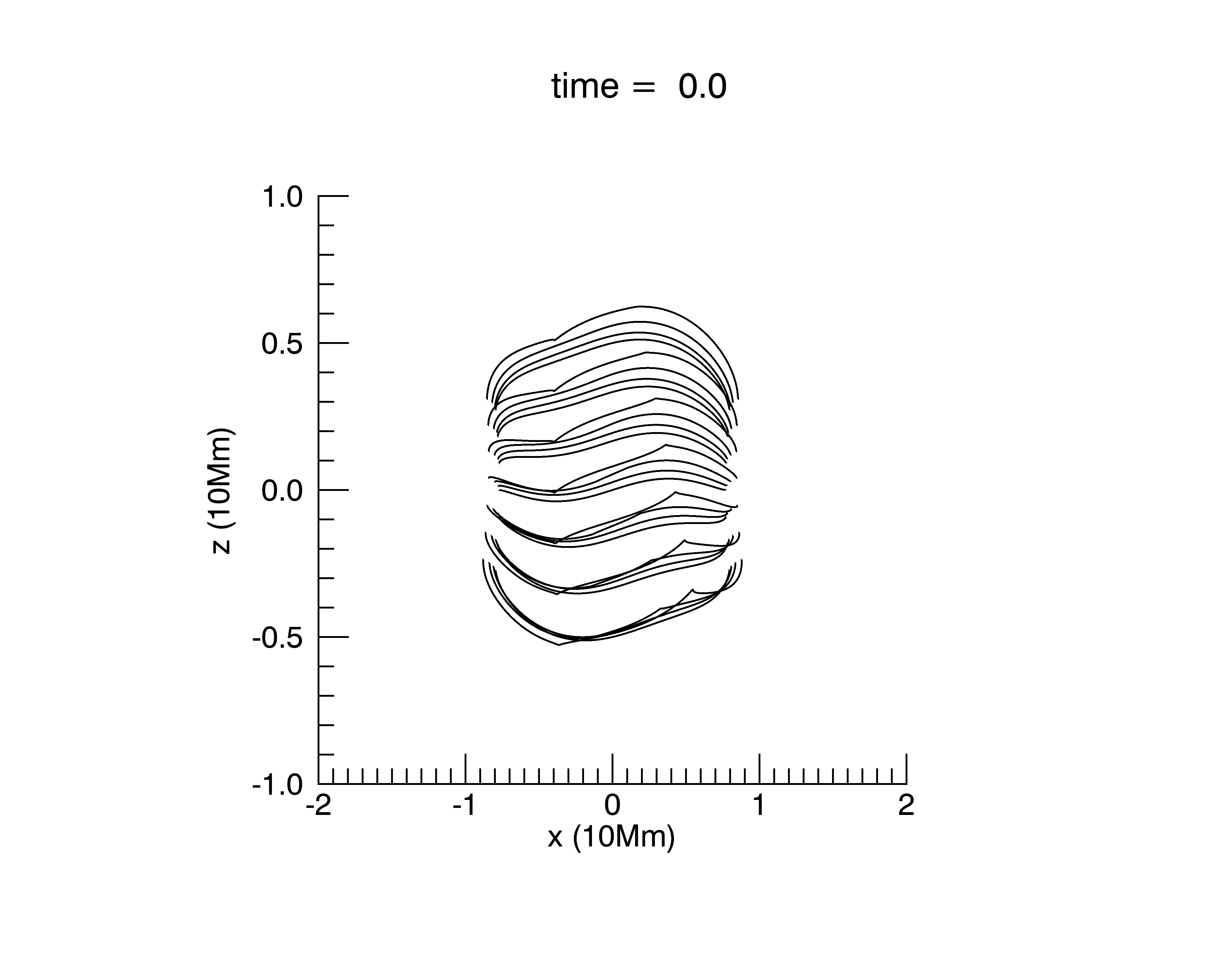}
    \includegraphics[width=\columnwidth, trim={1.5cm 1.5cm 1.5cm 1.5cm},clip]{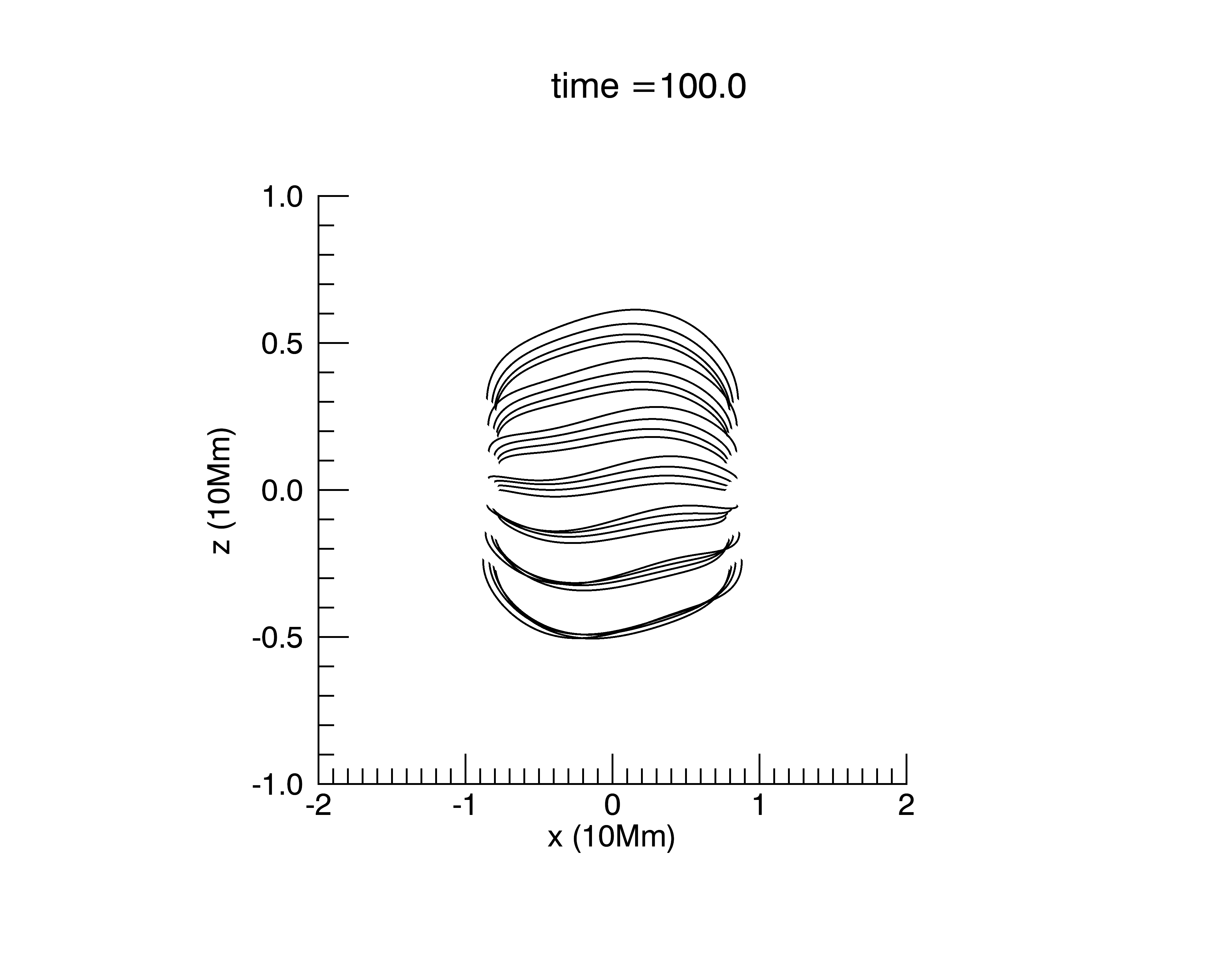}
    \caption{Projections of field lines from the 3D model detailed in Section \ref{ssec:CMTtheory} onto the $x-z$ plane showing the change in the field line shape between $t=0$ (top) and $t=100$ (bottom). Twist parameters are set as $\delta=1.0$, $a_{y}=1.0$ and $y_{h}=0.0$. Apparent sudden changes in the field line direction are caused by projection effects.}
    \label{fig:3dflinetwist}
\end{figure}

With a better understanding of how the downward collapse of field lines energises particles, we can now adjust some parameters in the model to better understand the impact of the field's untwisting on particle energisation. The most obvious parameter to work with is $\delta$, which, as seen in Equations (\ref{eqn:3dxinf}) and (\ref{eqn:3dzinf}), scales linearly with the strength of the twist
%
%
(see Figure~\ref{fig:3dflinetwist} for an example field line plot).
To investigate the impact of this parameter on particle energisation, we will take our regular 3D setup but with $\delta$ reduced from $1.0$ to $0.5$ and then to $0.0$, giving rise to a 3D configuration that is totally untwisted at all times.

\begin{figure}
    \centering
    \includegraphics[width=\columnwidth,trim={0.8cm 0.8cm 1.32cm 1.55cm},clip]{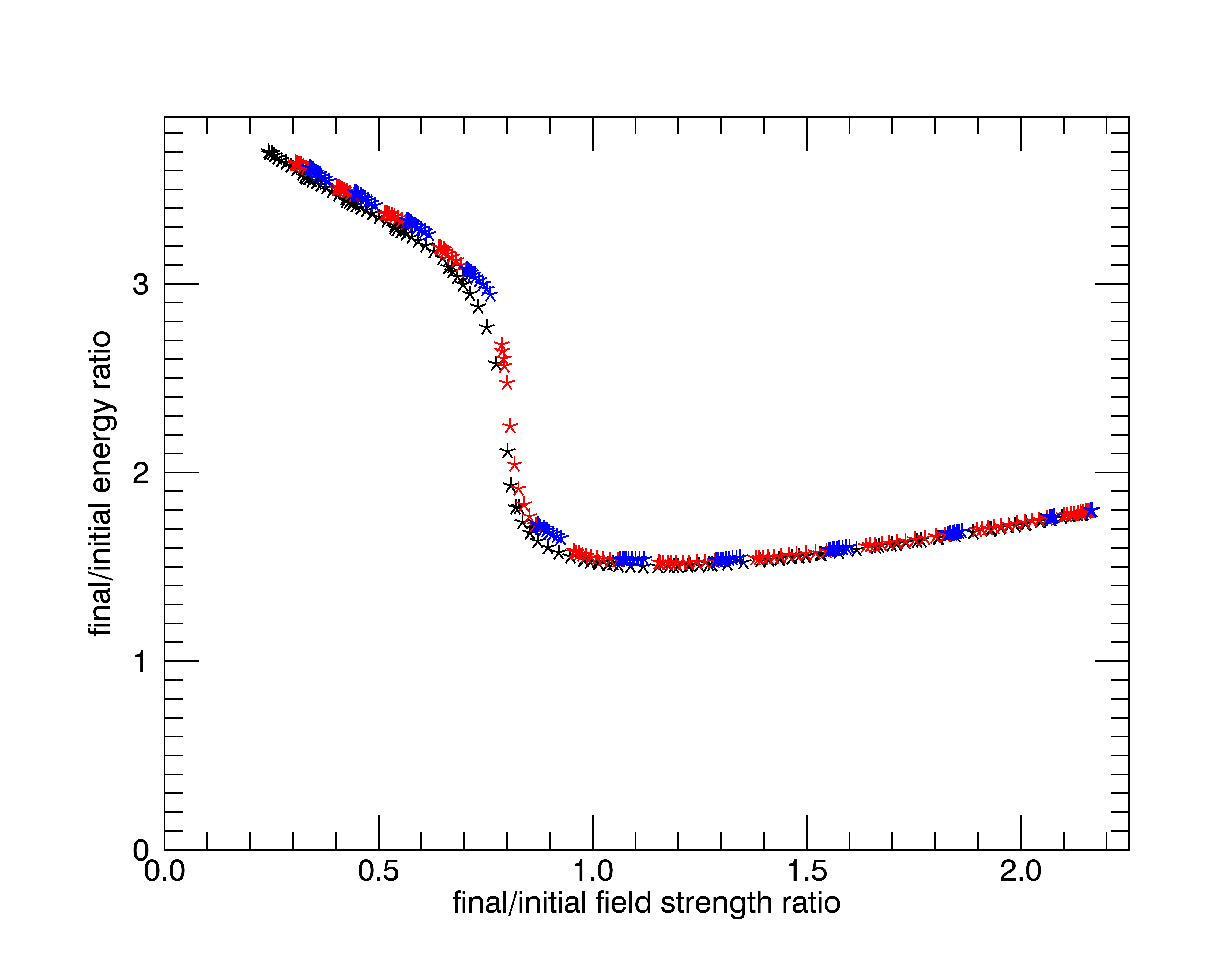}
    \caption{Energy ratio vs field ratio for the regular 3D standard case initial conditions with $\delta$ set to $1.0$ (black), $0.5$ (red) and $0.0$ (blue).}
    \label{fig:3ddeltacomp}
\end{figure}

We can see from Figure \ref{fig:3ddeltacomp} that varying $\delta$ does not have any significant effect on the shape of the energy curve. Most notably reducing $\delta$  causes points to gather in clusters. These clusters correspond to orbits starting at the same value for $x$, over a range of different $z$ values. It is unsurprising that points cluster in this way. Without twist the field will not vary as much in the $z$ direction, meaning that orbits starting at different $z$ values start on more characteristically similar field lines. The only contributions to $B_{z}$ will come from the magnetic source regions located at $z=0$. For the 3D setup without the twist, the grid of initial points at which orbits start needs to extend further in the $z$-direction so that particles start on field lines with different properties (curvature, spatial extent etc.).

A closer look at Figure \ref{fig:3ddeltacomp} shows that orbits in field configurations with smaller $\delta$ values gain slightly more energy through Fermi acceleration than those in more twisted fields. This contradicts general expectations of how the evolution of the field configuration would energise particles. One might expect that an initially twisted field which gradually untwists would energise particles more efficiently than an untwisted field, since this high-energy and non-potential field will relax towards the lower energy untwisted configuration. This raises the question of what effect the untwisting has on Fermi/betatron acceleration. 
Figure \ref{fig:3ddeltacomp} indicates that twisting the field may slightly reduce energy gains resulting from the collapse of field lines. One possibility is that the introduction of a $B_{z}$ component at the loop tops slightly flattens the local field, reducing Fermi acceleration. It is also possible that when adjusting the $\delta$ parameter, comparisons between particles on orbits starting at the same positions become less valid, as they may lie on field lines with more distinct properties. A more valid comparison may involve starting orbits lower in the loop legs where the twisting of the field is much reduced. This would see orbits start on comparable field lines but particles would be far more likely to escape the CMT before the end of the simulation, even for initial pitch angles close to $90^{\circ}$.

If, instead of looking at the curves that data points lie on, we look at individual initial positions for particle orbits, we see a quite different effect. Examining the orbits starting at $(-0.5,1.25,-0.5)$ (the particles with the lowest final/initial field strength ratio), the particle trajectory in the field with the strongest twist is associated with the greatest energy gain, despite being the trajectory that sees the smallest increase in magnetic field strength. This suggests that for certain orbits, the twist in the field reduces betatron acceleration, but that this is more than made up for by an increase in Fermi acceleration. A similar pattern can be observed for other orbits represented in the top left region of the plot (between 0.2 and 0.7 on the horizontal axis and 3.0 and 3.8 on the vertical axis). The reduced energy gains due to betatron acceleration may be the result of a larger initial field strength in the twisted field, as the twist strengthens the $B_{z}$ component. This reduces the magnetic moment and consequently, the energising effect of betatron acceleration. When comparing orbits with the same initial conditions in the region where betatron acceleration is dominant, particle trajectories starting further out in $z$ experience smaller overall energy gains because of a reduction in energisation due to betatron acceleration. 
Particles with higher initial pitch angles gain more energy due to betatron acceleration as they have a higher magnetic moment and particles with lower initial pitch angles gain more energy through Fermi acceleration as they take a larger $u_{\parallel}$ value at the loop top. If, as the results above suggest, particles with identical initial conditions gain more energy due to Fermi acceleration and less due to betatron acceleration in a more twisted field, then a twisted field may more efficiently energise particles with lower pitch angles and less efficiently energise particles with higher initial pitch angles. However, we would need to 
%
%
investigate a significant larger number of
particle orbits in order to test this hypothesis.

The blue points in Figure \ref{fig:3ddeltacomp} are too 
clustered to show the `transitional region' 
but when comparing the red points 
(corresponding to $\delta = 0.5$) and 
the black points (corresponding to $\delta = 1.0$), 
we see that the `transitional region' is notably steeper 
for the case with a lower $\delta$ value. 
This indicates that stronger twists make 
the minimum 
%
%
field strength
in the region swept out by a field line 
take 
%
%
on
a larger value due to the addition of the $B_z$ component. 

Another key feature of the 3D model is the 
%
%
%
decrease 
of the 
%
%
%
field line twist
with height. There are two parameters which govern the reduction of the twist with increasing height. Referenced in Equations (\ref{eqn:3dxinf}) and (\ref{eqn:3dzinf}), $a_{y}$ decides how strong the 
%
%
%
drop-off
is and $y_{h}$ decides the height at which the 
%
%
%
drop-off
first becomes significant. To investigate the effects of these parameters, we will take the same initial conditions as were used in Section \ref{sec:3Dresults} and first reduce $a_{y}$ from $1.0$ to $0.5$ and then we will increase $y_{h}$ from $0.0$ to $1.0$ (with $a_{y}$ returned to its initial value of $1.0$). 

\begin{figure}
    \centering
    \includegraphics[width=\columnwidth,trim={0.8cm 0.8cm 1.32cm 1.55cm},clip]{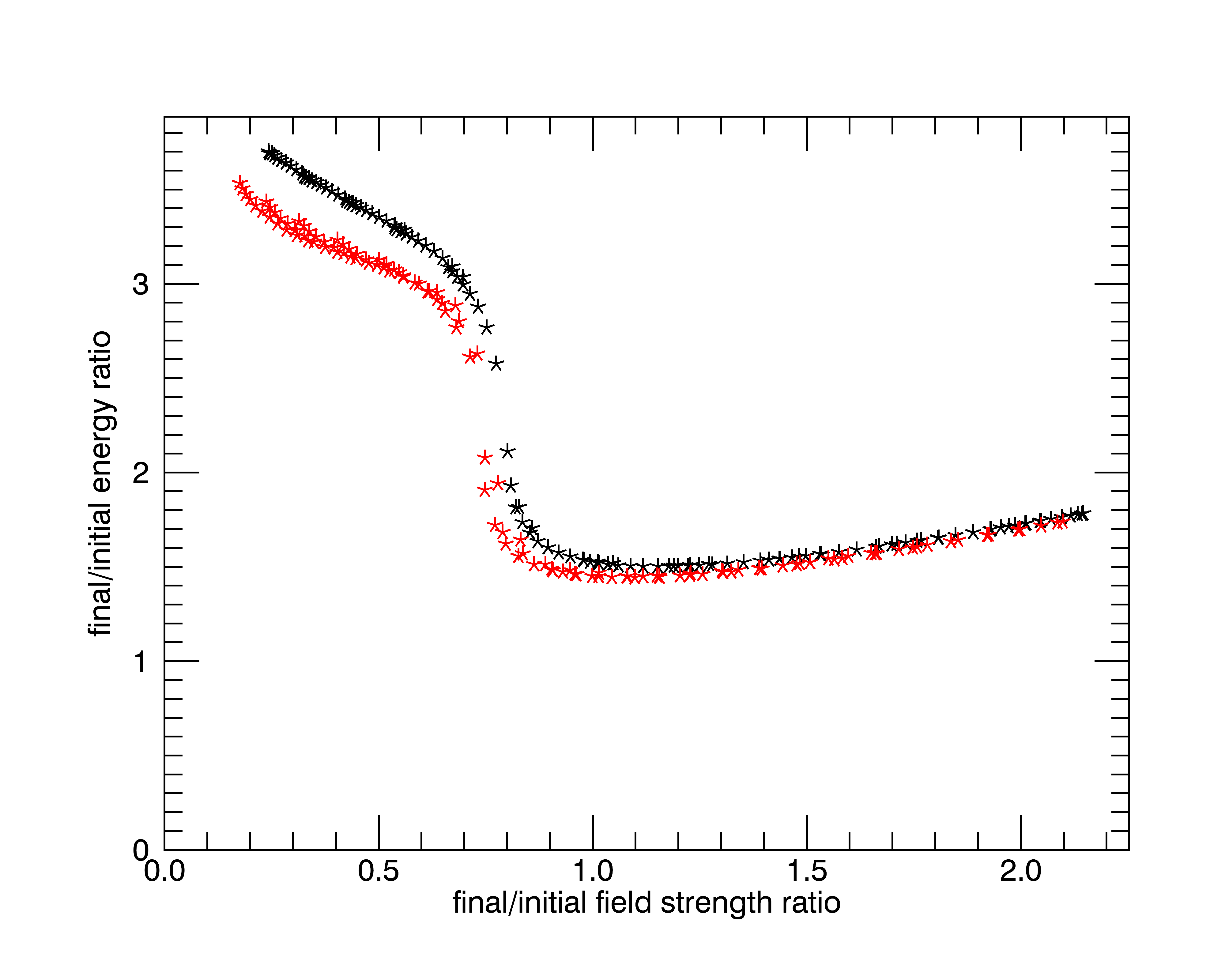}
    \caption{Energy ratio vs field ratio for the standard case initial conditions for the 3D model with $a_{y}=1.0$ (black) and $a_{y}=0.5$ (red).}
    \label{fig:3daycomp}
\end{figure}

Figure \ref{fig:3daycomp} presents the 
%
%
%
results
for $a_{y}$ = 1.0 and $a_{y}$ = 0.5. 
%
%
One immediately notices an overall reduction in energies with reduced $a_{y}$. Closer inspection of the orbits shows a reduction in both betatron and Fermi acceleration. 
The reduction in betatron acceleration is likely due to the stronger twist at the initial positions giving rise to a higher initial $B_{z}$ value. 
This reduces the value of 
%
%
the magnetic moment
$\mu$ and therefore the amount of energy that can be gained through betatron acceleration. The reduction in Fermi acceleration is more difficult to explain; it may be due to a flattening of loop tops for smaller $a_y$ values, as a reduced $a_y$ will lead to stronger $B_z$ higher up in the field. 

\begin{figure}
    \centering
    \includegraphics[width=\columnwidth,trim={0.8cm 0.8cm 1.32cm 1.55cm},clip]{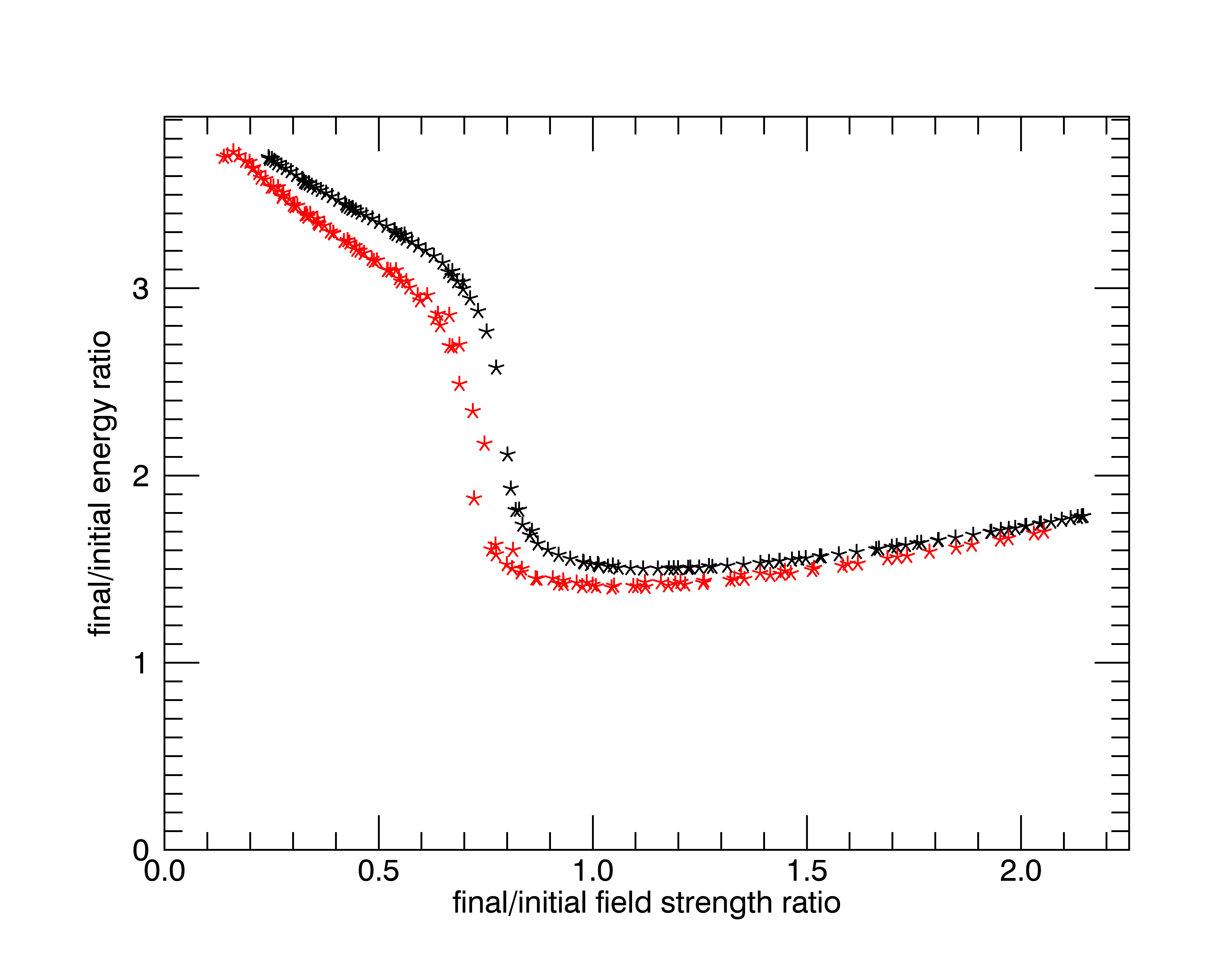}
    \caption{Energy ratio vs field ratio for the 3D model with $y_{h}=0.0$ (black) and $y_{h}=1.0$ (red). Particles are initialised with the conditions given in Section \ref{ssec:3Dparameters}}
    \label{fig:3dyhcomp}
\end{figure}

Figure \ref{fig:3dyhcomp} compares the 
%
%
%
cases where $y_{h}=0.0$ (black), in which the reduction in twist with height takes effect immediately and $y_h = 1.0$ (red), where this reduction only becomes significant above $y = 1.0$. It shows a decrease in the energy gained due to betatron acceleration, with red points being found to the left of corresponding black points. Because of this, orbits starting in the region where betatron acceleration is dominant experience lower energy gains in the field configuration with $y_{h}=1.0$. It is harder to assess how much of a difference there is in energy gain due to Fermi acceleration. Whilst corresponding orbits in the top left region (here located between 0.1 and 0.6 on the horizontal axis and 2.8 and 3.8 on the vertical axis) are separated horizontally, their final energies are quite similar. Points corresponding to the case with $y_h = 1.0$ show a smaller increase in the magnetic field so will gain less energy due to betatron acceleration. However, they show broadly the same energies as points starting with the same initial position in the case with $y_h = 0.0$, suggesting that they make up for this with increased Fermi acceleration. If we instead consider vertically aligned points, representing particles where the expected gains due to betatron acceleration are similar, we see that orbits gain more energy when $y_h = 0.0$, suggesting that Fermi acceleration is stronger in this configuration. This apparent contradiction shows that a more systematic test, involving a greater number of particles, is needed. 

Overall, these 
%
%
%
results do
not clearly explain the role of the untwisting of the field in particle energisation for 3D CMT models. Adjusting $\delta$ showed that particles gain less energy due to betatron acceleration in a more twisted field, with a greater contribution coming from Fermi acceleration. The reduction in energy gain resulting from betatron acceleration is shown consistently when $a_{y}$ and $y_{h}$ are adjusted to increase the twist in the field but there is conflicting evidence in how these parameters affect energy gains due to Fermi acceleration.

\section{Discussion and Conclusions}

In this paper we have investigated the specific processes responsible for particle energisation in 2D, 2.5D and 3D CMT models, finding that betatron acceleration is dominant for particles with orbits starting on collapsed field lines and that Fermi acceleration plays a greater role for orbits that start on the more stretched field lines. The curvature of collapsing loop tops is responsible for such Fermi acceleration 
\citep[e.g.][]{giuliani:etal05,eradat_oskoui:etal14}, as particles are accelerated by the interaction of the curvature drift with the electric field generated by the time-dependence of the magnetic field. In general, we found energy gains were modest for the majority of initial conditions. For the initial conditions tested in this paper most energies fall between 1.5 and 2.5 times the initial energy in the 2D and 2.5D models and between 1.5 and 3.5 times the initial energy in 3D. 

Previous work 
by
\citet{grady:etal12} 
for 2D CMTs similar to those investigated in this paper
has shown particle energy increases of factors of up to 40 for orbits with high initial pitch angles starting at the centre of the 2D CMT. We consider orbits with a far greater range of initial positions, which as discussed are all energised but only to a moderate degree. 
%
%
%
Orbits that see the greatest energy gains in a CMT only represent a small subset of initial conditions. Hence, the energisation of particles presented in this paper 
may 
not be efficient enough to explain observations of flare activity. 
%
%
However, to be able to properly assess the overall efficiency of CMT models for particle energisation
one would need to determine the time evolution
of a particle distribution function.
For this one has to
combine the calculation of
test particle orbits with a distribution
function of initial conditions to give
the appropriate probability weighting to different
initial conditions. To carry out such a calculation
is beyond the scope of this paper, but would
be useful to investigate whether, for example,
the CMT energisation process results in power law
distribution functions \citep[for an example using
a very simple CMT model, see e.g.][]{bogachev:somov07}.

One advantage of CMTs is that they can energise particle populations over large volumes. In this paper we see some evidence of this as all particles tested gain energy. 
Additionally, particles lying on the most stretched field lines, which when taken together cover a large volume, saw the strongest energy gains, reaching between 2.5 and 4 times their initial energy. This establishes our CMT model as likely being able to explain energisation of particles over a large volume. The models used in this paper have not been actively optimised to efficiently energise particles over large volumes, raising the possibility of further improvements in this direction.

We established that the role of energisation processes depends strongly on the initial position of the orbit. 
In particular, we observed a `transitional region', where small changes to the initial position of an orbit lead to a big difference in the energy gained due to Fermi acceleration. For the 2D and 2.5D models, this `transitional region' arises due to the substantial difference in field line length for field lines that initially lie either above or below the minimum in the field strength for the CMT field. In the 3D model a similar `transitional region' is present. Larger energy gains due to Fermi acceleration are associated with orbits starting on field lines which pass through the local minimum in the field strength as the field line relaxes. 

Speeding up the collapse of the 2D model shows that energy gains resulting from Fermi acceleration relate more closely to the total distance that a loop top travels during collapse, rather than the speed of collapse. This result may impact construction of more detailed CMT models. Particles are accelerated due to the betatron effect when a time- and space-dependent field strengthens and particles experience a stronger region of field. For models that assume ideal conditions, the key energisation processes of betatron and Fermi acceleration are determined by the time evolution of the magnetic field on a macroscopic scale. With this in mind, we anticipate the importance of these processes when constructing more realistic magnetic field models. 

We have presented 
results
that more thoroughly investigate the impact of twisting the initial configuration of field lines as first presented in 
\citet{grady:neukirch09}.
From our 
results, 
any differences in energisation due to untwisting are of minor importance compared to energisation resulting from the collapse of field lines. In each case field line collapse remains the key process in energising particles and in generating the features observed in energy ratio vs field strength ratio plots. 
It is not immediately clear whether the twist leads to greater or reduced particle energisation. Looking to Figure \ref{fig:3dyhcomp}, energy ratios for particles in more twisted field configurations 
are generally smaller than for particles in the untwisted field. 
However, inspection of individual orbits with identical initial conditions shows that particles in the twisted field experience lower energy gains relating to betatron acceleration offset by an increase in Fermi acceleration. This indicates that orbits starting in identical positions experience greater particle energy gains in the more twisted configuration. It therefore remains unclear what impact untwisting magnetic field has on particle acceleration. 

By only working with a selected group of initial conditions for particle orbits, we are unable to 
quantify global distributions and energies in CMTs. Further work 
is needed to understand the global motion of particles and evolution of 
particle energies, for example by moving beyond individual test particle orbits and modelling the evolution 
of a particle distribution function. An intermediate step, 
would consider a wider range of initial conditions for test particle 
orbits, paying particular attention to the relative number of trapped and escaping orbits. 
Our chosen initial conditions ensured 
particle trapping for the full duration of the simulation in order to better understand the 
energisation processes affecting particles that spent more time in the CMT. If sufficiently energised these 
particles
%
%
%
%
could 
potentially
correspond to the loop top X-ray source. 
%
%
A wider range of initial pitch angles may provide an estimate of the number of 
orbits that escape the CMT and detail the energisation processes impacting these 
particles. In our 
investigation, 
the energy gains due to Fermi 
acceleration generally depend on the velocity of the particle parallel to the field 
line and the distance covered by the loop top of the field line that the orbit lies 
on. A particle orbit with a low enough initial pitch angle to escape of the CMT early 
will have a greater contribution to particle energisation due to Fermi 
acceleration resulting from the higher parallel velocity of the particle at the loop top. 
This will be offset by the reduced distance that the loop top collapses through before the time of 
early escape. As a result, energy gains for particles that escape the trap are likely  
depend on both the escape time of the orbit and the starting 
position. In particular, we anticipate lower energy gains for particles on orbits 
that escape the CMT before their corresponding field lines have passed through the minimum in 
the field strength. 

%
%
Based on our kinematic models, Fermi acceleration can make an important contribution to
particle energisation in CMTs. Typical energy gains are between 0.5 and 2.5 times the initial particle energy. This raises the question of whether there are any 
aspects that our relatively simple models are missing which would increase energy gains. One such aspect could
be the presence of a braking region of the reconnection jet, which in extreme cases might
even lead to a termination shock \citep[e.g][]{miteva:mann07, mann:etal09}. In
the magnetospheric counterpart of CMTs, such a braking jet region has been strongly linked with 
accelerating particles via the Fermi process \citep[e.g][]{khotyaintsev-et-al2011, fu-et-al2011,artemyev14}. \citet{borissov:etal16} incorporated a braking jet into a 2D CMT model. This model did not, however, include a guide field and it has yet to be generalised to 3D, which would be a worthwhile future step. As the plasma dynamics inside 
a CMT are likely to be turbulent, another important aspect to consider for future CMT models would be the inclusion of stochastic scattering of the particle orbits, which may lead to changes in the energisation process as well as the conditions under which particle orbits remain trapped in the CMT or escape.






\section*{Acknowledgements}
The authors would like to thank the anonymous referee for useful and constructive comments that helped to improve the paper.
KM acknowledges financial support by the Science and Technology Facilities Council (STFC) through a PhD studentship (DTP ST/V507076/1) and TN acknowledges funding by
STFC Consolidated Grants ST/S000402/1 and ST/W001195/1.

\section*{Data Availability}

%
The data underlying the figures in this paper are available at https://doi.org/10.17630/dd7f39b7-d519-46bd-a6c2-f6dd2090a45f. The numerical code used to generate the data is available at https://github.com/KateMowbray/CMT\_particle\_orbit\_code.
 



\bibliographystyle{mnras}
\bibliography{TN-Bib-Files/particleacc-general,
TN-Bib-Files/TNpapers.bib,TN-Bib-Files/textbooks, TN-Bib-Files/CMT-refs} 








\bsp	
\label{lastpage}
\end{document}